\documentclass[
reprint,
superscriptaddress,
preprintnumbers,
nobibnotes,
nofootinbib,
amsmath,
amssymb,
aps,
floatfix,
longbibliography
]{revtex4-2}

\usepackage{gensymb}
\usepackage{mathrsfs}
\usepackage{graphicx}
\usepackage{dcolumn}
\usepackage{bm}
\usepackage{tabularx}
\usepackage{floatrow}
\usepackage{titlesec}
\usepackage{cancel}
\usepackage{amsmath}
\titleformat*{\section}{\normalsize\bfseries\filcenter}

\titlespacing*{\section}
{0pt}{3.0ex}{1.5ex}

\titlespacing*{\subsection}
{0pt}{2.5ex}{0.5ex}
\titlespacing*{\subsubsection}
{0pt}{2.0ex}{0.5ex}

\usepackage[
  colorlinks=true,
  urlcolor=blue,
  linkcolor=blue,
  citecolor=blue
]{hyperref}
\usepackage{xcolor}
\usepackage{textcomp}
\usepackage{amsmath}
\usepackage{physics}
\usepackage{siunitx}
\usepackage[artemisia]{textgreek}
\usepackage[normalem]{ulem}
\usepackage{upgreek}
\usepackage{etoolbox}

\usepackage{subfigure}
\usepackage{amsmath}
\usepackage{psfrag}
\usepackage{epsfig}
\usepackage{lineno}
\usepackage{hyperref}
\usepackage{amssymb}
\usepackage{notoccite}
\usepackage{url}
\usepackage{xcolor}
\usepackage[colorinlistoftodos]{todonotes} 
\usepackage{verbatim}

\ProvideTextCommandDefault{\textonehalf}{${}^1\!/\!{}_2\ $}

\begin{document}
\title{Time-resolved diamond magnetic microscopy of superparamagnetic iron-oxide nanoparticles}

\author{B.~A.~Richards$^{\mathsection}$}
\affiliation{Center for High Technology Materials, 
University of New Mexico, Albuquerque, NM, USA}
\affiliation{Department of Physics and Astronomy,
University of New Mexico, Albuquerque, NM, USA}

\author{N.~Ristoff$^{\mathsection}$}
\affiliation{Center for High Technology Materials, 
University of New Mexico, Albuquerque, NM, USA}
\affiliation{Department of Physics and Astronomy,
University of New Mexico, Albuquerque, NM, USA}

\author{J.~Smits}
\affiliation{Center for High Technology Materials, 
University of New Mexico, Albuquerque, NM, USA}

\author{A.~Jeronimo Perez}
\affiliation{Center for High Technology Materials, 
University of New Mexico, Albuquerque, NM, USA}

\author{I.~Fescenko}
\affiliation{Center for High Technology Materials, 
University of New Mexico, Albuquerque, NM, USA}
\affiliation{Laser Center of the University of Latvia, Riga, Latvia}

\author{M.~D.~Aiello}
\affiliation{Center for High Technology Materials, 
University of New Mexico, Albuquerque, NM, USA}
\affiliation{Department of Physics and Astronomy,
University of New Mexico, Albuquerque, NM, USA}

\author{F.~Hubert}
\affiliation{Center for High Technology Materials, 
University of New Mexico, Albuquerque, NM, USA}
\affiliation{Department of Physics and Astronomy,
University of New Mexico, Albuquerque, NM, USA}

\author{Y.~Silani}
\affiliation{Center for High Technology Materials, 
University of New Mexico, Albuquerque, NM, USA}
\affiliation{Department of Physics and Astronomy,
University of New Mexico, Albuquerque, NM, USA}

\author{N.~Mosavian}
\affiliation{Center for High Technology Materials, 
University of New Mexico, Albuquerque, NM, USA}
\affiliation{Department of Physics and Astronomy,
University of New Mexico, Albuquerque, NM, USA}

\author{M.~Saleh~Ziabari}
\affiliation{Center for High Technology Materials, 
University of New Mexico, Albuquerque, NM, USA}
\affiliation{Department of Physics and Astronomy,
University of New Mexico, Albuquerque, NM, USA}
\affiliation{Center for Integrated Nanotechnologies, Sandia National Labs, Albuquerque, NM, USA}

\author{A.~Berzins}
\affiliation{Center for High Technology Materials, 
University of New Mexico, Albuquerque, NM, USA}

\author{J.~T.~Damron}
\affiliation{Center for High Technology Materials, 
University of New Mexico, Albuquerque, NM, USA}
\affiliation{Oak Ridge National Laboratory, Oak Ridge, TN, USA}

\author{P.~Kehayias}
\affiliation{Center for High Technology Materials, University of New Mexico, Albuquerque, NM, USA}
\affiliation{Sandia National Labs, Albuquerque, NM, USA}

\author{D.~Egbebunmi}
\affiliation{Department of Mechanical and Materials Engineering, University of Nebraska, Lincoln, NE, USA}

\author{J.~E.~Shield}
\affiliation{Department of Mechanical and Materials Engineering, University of Nebraska, Lincoln, NE, USA}

\author{D.~L.~Huber}
\affiliation{Center for Integrated Nanotechnologies, Sandia National Labs, Albuquerque, NM, USA}

\author{A.~M.~Mounce}
\affiliation{Center for Integrated Nanotechnologies, Sandia National Labs, Albuquerque, NM, USA}

\author{M.~P.~Lilly}
\affiliation{Center for Integrated Nanotechnologies, Sandia National Labs, Albuquerque, NM, USA}

\author{T.~Karaulanov}
\affiliation{Predigma Inc., Boulder, CO, USA}

\author{A.~Jarmola}
\affiliation{ODMR Technologies Inc., El Cerrito, CA, USA}
\affiliation{Department of Physics, University of California, Berkeley, CA, USA}

\author{A.~Laraoui}
\affiliation{Center for High Technology Materials, 
University of New Mexico, Albuquerque, NM, USA}
\affiliation{Department of Mechanical and Materials Engineering, University of Nebraska, Lincoln, NE, USA}
\affiliation{Department of Physics and Astronomy, Nebraska Center for Materials and Nanoscience, University of Nebraska, Lincoln, NE, USA}

\author{V.~M.~Acosta}
\email[Email address:~]{victormarcelacosta@gmail.com}
\affiliation{Center for High Technology Materials, 
University of New Mexico, Albuquerque, NM, USA}
\affiliation{Department of Physics and Astronomy,
University of New Mexico, Albuquerque, NM, USA}

\renewcommand{\thefootnote}{}{\footnote{\vspace{0.5mm}\hspace{-2.5mm}$\mathsection$ B.~A.~Richards and N.~Ristoff  contributed equally to this work.}}

\date{\today}

\begin{abstract}
Superparamagnetic iron-oxide nanoparticles (SPIONs) are promising probes for biomedical imaging, but the heterogeneity of their magnetic properties is difficult to characterize with existing methods. Here, we perform widefield imaging of the stray magnetic fields produced by hundreds of isolated ${\sim}30$-nm SPIONs using a magnetic microscope based on nitrogen-vacancy centers in diamond. By analyzing the SPION magnetic field patterns as a function of applied magnetic field, we observe substantial field-dependent transverse magnetization components that are typically obscured with ensemble characterization methods. We find negligible hysteresis in each of the three magnetization components for nearly all SPIONs in our sample. Most SPIONs exhibit a sharp Langevin saturation curve, enumerated by a characteristic polarizing applied field, $B_c$. The $B_c$ distribution is highly asymmetric, with a standard deviation ($\sigma_c=1.4~{\rm mT}$) that is larger than the median ($0.6~{\rm mT}$). Using time-resolved magnetic microscopy, we directly record SPION Néel relaxation, after switching off a $31~{\rm mT}$ applied field, with a temporal resolution of ${\sim}60~{\rm ms}$ that is limited by the ring-down time of the electromagnet coils. For small bias fields $|B_{\rm hold}|=1.5{\mbox{-}}3.5~{\rm mT}$, we observe a broad range of SPION Néel relaxation times--from milliseconds to seconds--that are consistent with an exponential dependence on $B_{\rm hold}$. Our time-resolved diamond magnetic microscopy study reveals rich SPION sample heterogeneity and may be extended to other fundamental studies of nanomagnetism.
\end{abstract}

\maketitle
\twocolumngrid

\begin{figure*}[htb]
    \includegraphics[width=\textwidth]{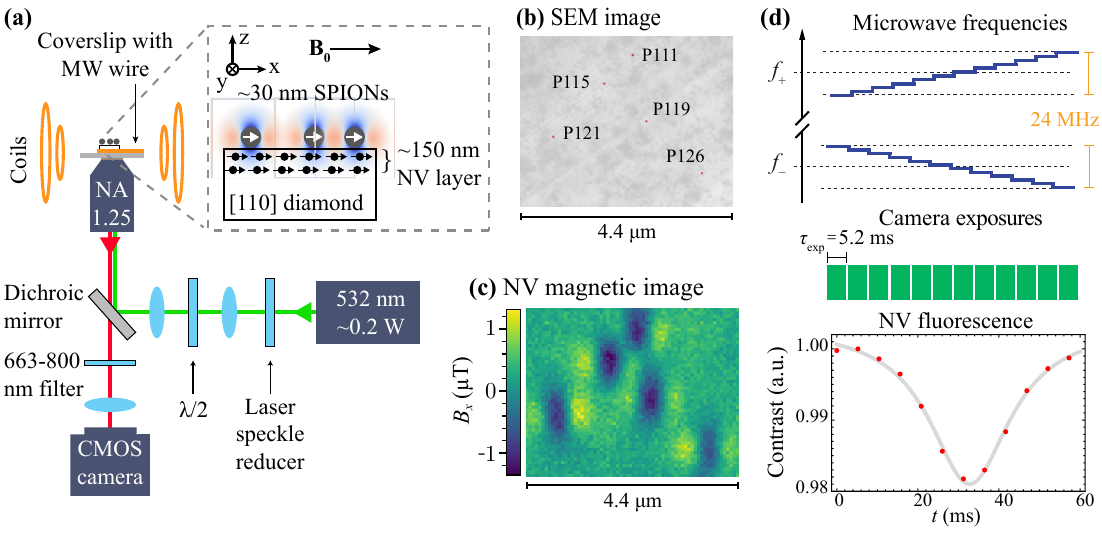}
    \caption{\textbf{Experimental setup.} (a) Layout of the apparatus. Light from a $532~{\rm nm}$ laser passes through an oscillating laser speckle reducer, followed by a half-waveplate. The beam is focused to a ${\sim}45~{\rm \upmu m}$ diameter on the diamond via lenses and a microscope objective ($100\times$ magnification, 1.25 numerical aperture). NV fluorescence is collected through the same objective, spectrally filtered, and imaged onto a scientific CMOS image sensor. Two pairs of electromagnet coils are used to apply the magnetic field. The inset shows the diamond chip, NV layer, and SPIONs. (b) SEM image of a portion of the FOV with five isolated SPIONs on the diamond surface, highlighted in red (\ref{app:SEM}). (c) Diamond magnetic microscope image of the FOV sub-region in (b) at $B_0=30~{\rm mT}$. (d) Magnetic imaging timing diagram. Microwave frequencies (blue) are swept simultaneously about both NV $f_{\pm}$ spin transitions. The $5.2~{\rm ms}$ camera exposures capture NV fluorescence, synchronized with the microwave frequency steps. The intensity of an example pixel is plotted as a function of time (alternatively: microwave detuning) and fit to a Lorentzian curve. The central frequency offset is proportional to the magnetic field component along the NV axis due to the sample, $B_x$.}
    \label{fig:schematic}
\end{figure*}

\section{Introduction}
\label{sec:introduction}
Superparamagnetic iron oxide nanoparticles (SPIONs) are promising probes for biomedical applications because they are relatively non-toxic~\cite{RED2012}, their magnetic fields are not attenuated by tissue, and they can be controlled spatially and temporally \textit{in vivo}~\cite{DOB2008}. They are widely investigated for use in cancer therapy, such as hyperthermia~\cite{SIL2011,SUC2020,SAM2021,VAN2023}, and in diagnostic imaging, including magnetic resonance imaging~\cite{EST2015}, magnetic particle imaging~\cite{PAN2015}, and magneto-relaxometry~\cite{JOH2012}. Many of these applications would benefit from using SPIONs with well-characterized, uniform magnetic properties. 

Measurement of the distributions of nanoparticle magnetic properties is needed for manufacturing quality control~\cite{STA2018}. This is especially important for SPIONs, as their Néel relaxation (spontaneous flipping of internal magnetization) time is an exponential function of magnetic core volume \cite{NEE1953, BRO1963}, leading to high heterogeneity in magnetic dynamics even in relatively monodisperse samples. Most existing magnetic measurement instruments~\cite{JAU2021,TAY2016}, such as those based on superconducting quantum interference devices (SQUIDs)~\cite{BUC2018}, feature high throughput operation under variable conditions, but they typically have a detection threshold of millions of SPIONs, complicating inference of particle heterogeneity~\cite{TAU2012,BOG2015,TOP2019,SAA2021}. Methods based on the magneto-optical Kerr effect offer exquisite temporal resolution, down to ${\sim}100~{\rm fs}$~\cite{AND2006,LAR2007}, but they also lack single-nanoparticle-level sensitivity.

Magnetic measurements of single nanoparticles have been made using magnetic force microscopy~\cite{LED1994}, micro-SQUIDs~\cite{THI2003}, and transmission electron microscopes modified for Lorentz microscopy~\cite{HEF1991}. Each of these methods can resolve individual nanoparticles with core diameters $\lesssim30~{\rm nm}$, but they either lack high throughput, require cryogenic temperatures, and/or involve perturbative sample-tip interactions. A recent photothermal circular dichroism method offers faster, room-temperature measurements, but it requires calibration to infer quantitative information~\cite{ADH2024}.

Magnetic microscopy based on nitrogen-vacancy (NV) centers in diamond offers an alternative route to quantitatively characterize SPION magnetic properties under a variety of conditions. Diamond magnetic microscopy has been used to image magnetism in numerous nanoparticle samples including ferritin~\cite{WAN2019, NOL2014}, iron deposits in organelles~\cite{MCC2020, DEG2021}, malarial hemozoin crystals~\cite{FES2019}, spin crossover Fe-triazole nanorods~\cite{LAM2023}, and immuno-magnetically labeled cells~\cite{GLE2015, DAV2018, CHE2022}. The magnetic field of individual SPIONs has been imaged~\cite{GOU2014,MOS2024}, and other properties of iron-oxide particles have been observed~\cite{CAM2021, SAD2018, BAR2020, KUW2020, SMI2016,MAT2024}. The AC susceptibility of magnetic samples has also been imaged~\cite{ZHA2021, DAS2023}. However, imaging the magnetic dynamics of numerous individual SPIONs is a capability that remains to be demonstrated.

Here, we use diamond magnetic microscopy to image the field-dependent magnetization and time-domain Néel relaxation of over a hundred individual SPIONs in parallel. We compile histograms and extract statistics of key SPION parameters, such as saturation magnetic moment and Néel relaxation time, to characterize the ensemble distribution. The data reveal remarkable properties, including starkly steeper magnetization curves for some SPIONs compared to those obtained from ensemble methods. Such a rich heterogeneity is masked in ensemble methods, suggesting our method can produce complementary insights in fundamental studies of nanomagnetism.

\section{Experimental setup}
\label{sec:experimental setup}

For diamond magnetic microscopy of SPIONs, we used the experimental setup depicted in Fig.~\ref{fig:schematic}(a) (see also~\ref{app:setup}). The diamond is a $100\mbox{-}{\rm \upmu m}$-thick electronic-grade membrane with faces polished normal to a [110] crystal direction such that two of the four possible NV axes lie in plane. Ion implantation and annealing is used to form a ${\sim}150~{\rm nm}$-thick NV layer near the diamond surface on which SPION samples are dispersed (\ref{app:diamond}). A beam of 532-nm light continuously illuminates NV centers in a $45\times45~{\rm \upmu m}^2$ field of view (FOV) via an oil-immersion objective. NV center fluorescence is collected through the same objective, spectrally filtered (passing $663{\mbox{-}}800~{\rm nm}$), and imaged onto a scientific CMOS image sensor, with a lateral resolution of ${\sim}350~{\rm nm}$. The diamond is mounted on a glass coverslip using UV-curing adhesive. Microwaves are delivered to the NV centers via copper lines printed on the coverslip. The diamond is positioned so that one of the in-plane NV axes lies parallel to an applied magnetic field $\vec{B_0}$, generated by two pairs of coils. The diamond-coverslip assembly is mounted on a 3-axis piezo-actuated translation stage. Fluorescence images of fiducial markers milled into the diamond NV layer are used to adjust the translation stage and stabilize the diamond position~\cite{MOS2024}, limiting drift to ${\lesssim}100~{\rm nm/day}$.

The SPIONs (Ocean Nanotech SOR30) have Fe$_3$O$_4$ magnetic cores and an oleic acid coating. Through transmission electron microscopy (TEM) analysis (\ref{app:tem}), we found that the shape of the nanoparticles is predominantly spherical, with a mean diameter of $30~{\rm nm}$ and standard deviation of ${\sim}4~{\rm nm}$. Selected area electron diffraction patterns and lattice fringe spacings are consistent with the face-centered cubic Fe$_3$O$_4$ crystal structure. High-resolution TEM images also showed evidence that the particles are polycrystalline. 

The initial SPION suspensions are diluted in hexane, drop-cast onto the diamond, and dried in air, resulting in isolated single SPIONs with an average spacing of ${\sim}3~{\rm \upmu m}$~(\ref{app:SPION}). SPIONs lying within the diamond magnetic microscope's FOV are identified by scanning electron microscopy (SEM) and co-localized with respect to the fiducial markers. In the specific FOV presented here (\ref{app:SPION}), we identified and localized $259$ single SPIONs, of which $101$ were sufficiently isolated (${\gtrsim}1.5~{\rm \upmu m}$ from nearest neighbor and image edges) such that their magnetic dipole patterns could be individually resolved with negligible overlap in our diamond magnetic microscope. Figure~\ref{fig:schematic}(b) shows a SEM image of five isolated SPIONs on the diamond surface within a sub-region of this FOV.

Figure \ref{fig:schematic}(c) shows a diamond magnetic microscopy image of the same sub-region as in Fig.~\ref{fig:schematic}(b) under an applied field $B_0=30~{\rm mT}$. Five magnetic features are observed in the same locations as the SPION features in the SEM image. To construct a magnetic image, we use a dual-resonance optically detected magnetic resonance technique~\cite{GLE2015,WOJ2018A,FES2020,KAZ2021,SHI2022,YU2024}, Fig.~\ref{fig:schematic}(d) (see also~\ref{app:NV} and \ref{app:processing}). The frequencies of two microwave tones are simultaneously discretely swept in opposite directions about both NV spin resonances. The sweeps consist of $12$ frequency steps, with a dwell time of $5.2~{\rm ms}$, for a total sweep length of $62~{\rm ms}$. The sweeps of each tone are centered about different NV spin transition frequencies, $f_{\pm} \approx D \pm \gamma_{\rm nv} B_0$, where $D\approx2870~{\rm MHz}$ is the NV zero-field splitting and $\gamma_{\rm nv}=28.03~{\rm GHz/T}$ is the NV gyromagnetic ratio. Concurrently, the camera records a continuous series of frames, with exposure times synchronized to the microwave frequency steps. The directions of the microwave sweeps are periodically swapped and frames of the same microwave frequencies are averaged together to remove rolling-shutter and other timing artifacts~\cite{FES2019}. For a given camera pixel, the intensity as a function of microwave detuning is fit to a Lorentzian function. The fitted central frequency offset, $\Delta_c=\gamma_{\rm nv} B_x$, reveals the additional magnetic field component along the NV axis due to the sample, $B_x$~(\ref{app:NV}). Throughout, we define the $z$-axis to be normal to the diamond surface and the $x$-axis to be along the NV and $\vec{B_0}$ axes, see inset of Fig.~\ref{fig:schematic}(a). The fitting procedure is repeated for each pixel in the FOV in order to generate a magnetic image. Compared to sequential interrogation of the $f_{\pm}$ resonances~\cite{GLE2015,FES2019}, this simultaneous dual-resonance method provides an increase in fluorescence contrast, with minimal change in resonance linewidth, leading to a ${\sim}1.4$-fold improvement in sensitivity (\ref{app:dualsens}).

\begin{figure*}[hbt]
    \centering \includegraphics[width=\textwidth]{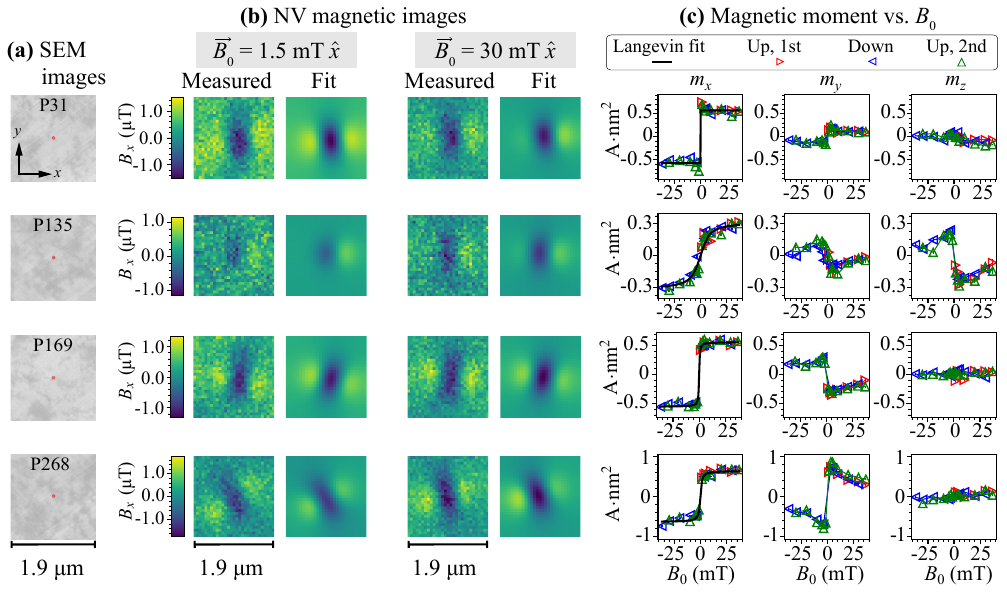}
    \caption{\textbf{Field-dependent magnetization of single SPIONs.} (a) SEM images of four example SPIONs in $1.9\times1.9~{\rm \upmu m^2}$ FOV sub-regions. SPIONs are highlighted in red (\ref{app:sample}). (b) Magnetic images (left) of the SPIONs  along with fits (right) at two example applied fields, $B_0=1.5~{\rm mT}$ and $30~{\rm mT}$. (c) Fitted values of magnetic-moment components $m_x, m_y, m_z$ versus $B_0$. Over the course of several days, $B_0$ is swept up, down, and back up, with negligible observable hysteresis. Error bars corresponding to fit uncertainty are typically smaller than the plot markers. The $m_x(B_0)$ curves are fit to Eq.~\eqref{eq:langevin}.} 
    \label{fig:case_study}
\end{figure*}

\begin{figure*}[htb]
    \centering
    \includegraphics[width=\textwidth]{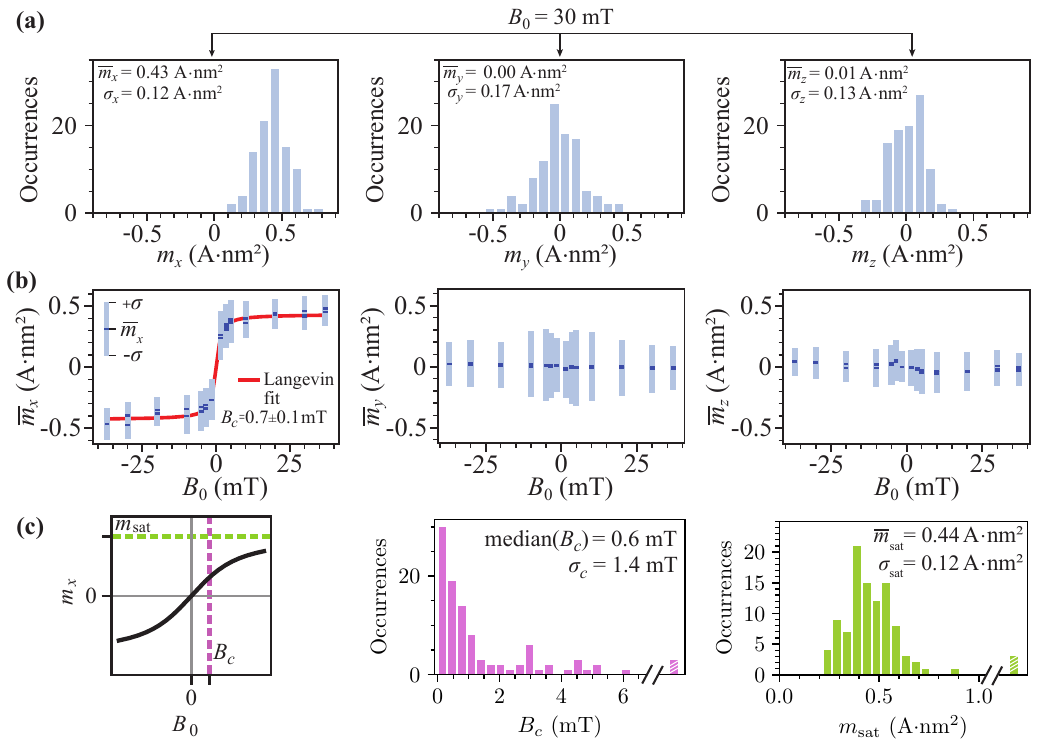}
    \caption{\textbf{Statistics for field-dependent magnetization.} a) Distributions of SPION magnetic-moment components for an applied field $B_0 = 30$ mT. b) Mean magnetic-moment components (from left to right: $\overline{m}_x, \overline{m}_y,~{\rm and}~\overline{m}_z$) as a function of $B_0$. The mean is taken over all fitted $m_x, m_y, m_z$ components for the sample of 101 single SPIONs. The $\overline{m}_x(B_0)$ data are fit to Eq.~\eqref{eq:langevin}. c) Histograms of single-SPION $m_x(B_0)$ Langevin fit parameters. (left) visual interpretation of parameters, (middle) characteristic field $B_c$, (right) saturation magnetization $m_{\rm sat}$. There are 3 SPIONs with $B_c$ and $m_{\rm sat}$ values that are outliers due to poor fits (SPIONs P109, P117, and P140, see \ref{app:data}) and are omitted from statistics calculations.}
    \label{fig:avg_lang}
\end{figure*}

\section{Field-dependent magnetic moments}
\label{sec:magnetization measurement}
We first studied the magnetization response of individual SPIONs as a function of $B_0$. Figure~\ref{fig:case_study} displays results for four example particles. An SEM image of each SPION is shown in Fig.~\ref{fig:case_study}(a). Figure \ref{fig:case_study}(b) shows magnetic images of each SPION at two of the applied fields, $B_0 = 1.5~{\rm mT}$ and $30~{\rm mT}$. Notably, the magnetic field patterns are different for each SPION, and they also vary based on $B_0$, indicating complex, field-dependent variations in the time-averaged SPION magnetic moment vector, $\vec{m}$. To determine $\vec{m}$, we fit each magnetic image to a function for the $x$-component of the magnetic field of a point dipole, averaged over a $150~{\rm nm}$-thick NV layer, and convolved with an effective Gaussian point-spread function (\ref{app:dipoleFits}). The fits are shown alongside the measured images in Fig.~\ref{fig:case_study}(b), showing good agreement. 

Figure~\ref{fig:case_study}(c) shows plots of the fitted magnetic-moment components as a function of $B_0$. For nearly every SPION, we do not observe evidence for hysteresis, at least over the days-long field sweeps studied here. The $m_x$ component is parallel to $B_0$, and it is expected to follow a Langevin saturation curve, similar to that observed in ensemble experiments~\cite{BEA1959}. We fit each SPION's $m_x(B_0)$ curve to the equation:
\begin{align}
\label{eq:langevin}
    m_x = m_{\rm{sat}}\left[ \coth\left(\frac{B_0}{B_c}\right) - \frac{B_c}{B_0} \right]
\end{align}
where $m_{\rm sat}$ is the saturation magnetic moment and $B_c$ is a characteristic polarizing applied field. An interpretation of these parameters is illustrated in Fig.~\ref{fig:avg_lang}(c). Figure~\ref{fig:case_study}(c) shows fits to Eq.~\eqref{eq:langevin} for four example SPIONs. Notably, we find step-like $m_x(B_0)$ curves for some SPIONs (for example, P31 and P169), indicating that their magnetization already saturates under very weak applied fields $B_c\ll1.5~{\rm mT}$. Other SPIONs, such as P135, exhibit more gradual saturation.

The $m_y$ and $m_z$ components are transverse to $\vec{B_0}$ and exhibit different behavior from particle to particle. However, they share some universal qualitative features that can be understood as follows. At large values of $|B_0|$, $\vec{m}$ tends to align along the $x$-axis, and thus $m_y$ and $m_z$ tend towards zero. For $B_0\approx0$, the SPION's instantaneous magnetic moment flips back and forth along its easy axis, on a timescale much shorter than the image acquisition, resulting in a time-averaged magnetic moment $\vec{m}\approx0$. However for intermediate values of $|B_0|$, the effective easy axis of each SPION partially rotates towards the magnetic field, and the occupation probability of the magnetic moment along each direction of the easy axis is no longer equal~\cite{COF1995,CHA1983}. The exact rotation angles and occupation probabilities depend on the anisotropy constant and the angle of the easy axis, which are different for every SPION. The combined effect results in non-zero components for $m_y$ and $m_z$ at intermediate applied fields, that vary from SPION to SPION.

We conducted a similar analysis on all $101$ isolated {SPIONs} in the FOV. Plots of the fitted magnetic-moment components versus $B_0$ for each SPION are shown in~\ref{app:data}. For a given value of $B_0$, we make histograms of $m_x$, $m_y$, and $m_z$ and extract statistics of the ensemble. Figure \ref{fig:avg_lang}(a) shows these histograms at $B_0=30~{\rm mT}$. At this field, we find $\overline{m}_x=0.43~{\rm A{\cdot} nm^2}$ with an ensemble standard deviation $\sigma_x=0.12~{\rm A{\cdot} nm^2}$. The mean values of $\overline{m}_y$ and $\overline{m}_z$ are approximately zero, as expected for randomly-aligned easy axes.

Figure~\ref{fig:avg_lang}(b) plots the mean values $\overline{m}_x$, $\overline{m}_y$, and $\overline{m}_z$ as a function of $B_0$, with error bars representing the ensemble standard deviations. The $\overline{m}_x(B_0)$ curve is fit to a Langevin function, Eq.~{\eqref{eq:langevin}, revealing $B_c = 0.7 \pm 0.1~{\rm mT}$ and $m_{\rm sat}=0.43 \pm 0.01~{\rm A{\cdot}nm^2}$, where the confidence intervals represent the fit uncertainty. The values of $\overline{m}_y$ and $\overline{m}_z$ are close to zero for all $B_0$, but the standard deviations show hints of a qualitative pattern of first increasing and then decreasing as $|B_0|$ increases. This is consistent with our previous discussion of an intermediate field regime where the transverse magnetization of individual SPIONs is maximal.

Figure~\ref{fig:avg_lang}(c) shows histograms of the fitted values of $B_c$ and $m_{\rm sat}$ for the $m_x(B_0)$ curves of each of the 101 SPIONs. The $m_{\rm sat}$ distribution resembles a normal distribution with mean $\overline{m}_{\rm sat}=0.44~{\rm A{\cdot} nm^2}$ and standard deviation $\sigma_{\rm sat}=0.12~{\rm A {\cdot} nm^2}$. The mean is consistent with, though on the lower end, of the range of values reported in the literature for similar SPIONs~\cite{AHR2015,WU2021,MOS2024,CHE2021} (\ref{app:bulk}, \ref{app:dipoleFits}). The lower $\overline{m}_{\rm sat}$ may be related to the polycrystalline structure~\cite{NED2017} observed by high-resolution TEM~(\ref{app:tem}). Notably, the $B_c$ distribution is highly asymmetric, with a peak near $B_c=0$ and a long tail. This reflects the presence of a large number of individual SPIONs with very sharp, step-like $m_x(B_0)$ curves. The ensemble-mean behavior shown in Fig.~\ref{fig:avg_lang}(b) is more consistent with the median of this distribution, ${\rm median}(B_c)=0.6~{\rm mT}$, whereas the standard deviation is larger, $\sigma_c=1.4~{\rm mT}$. 

\begin{figure*}[htb]
    \centering
    \includegraphics[width=\textwidth]{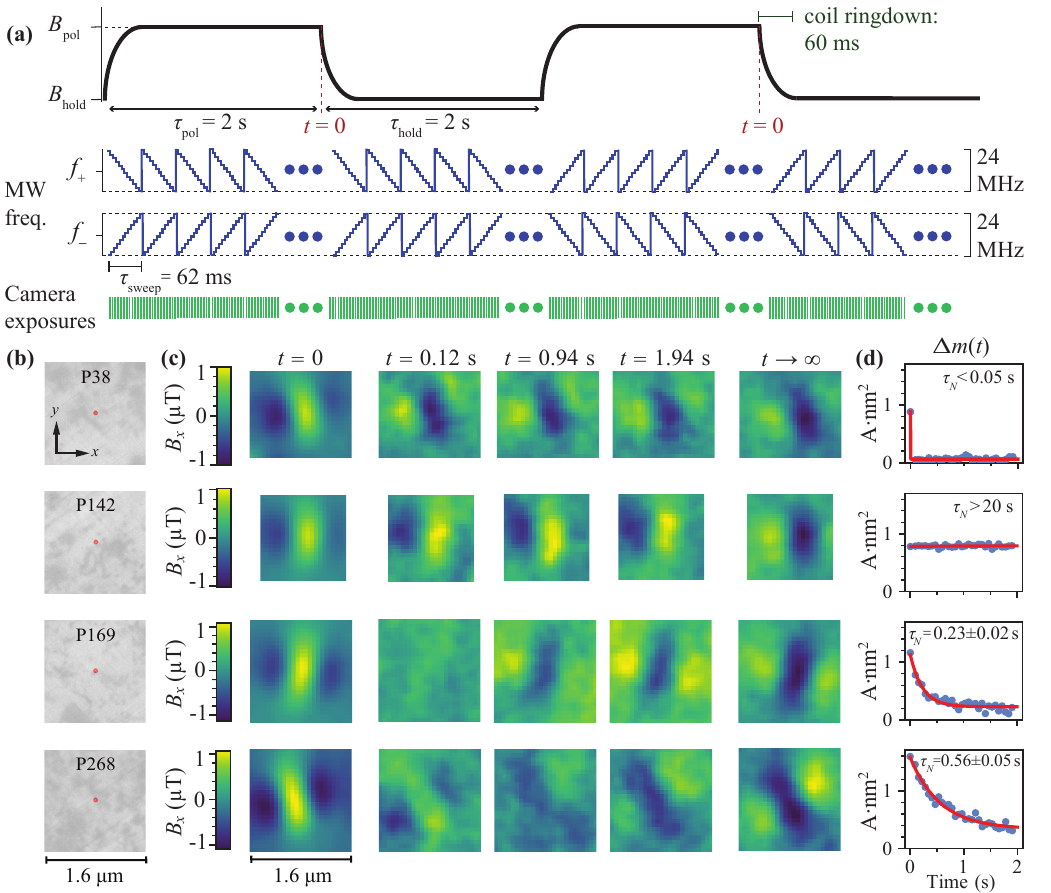} 
    \caption{\textbf{Time-resolved diamond magnetic microscopy of N{\`e}el relaxation.} (a) Measurement sequence depicting the switching of applied magnetic field, dual-resonance microwave frequency sweeping, and camera exposures. (b) SEM images of four single SPIONs. Each image is a $1.6{\times}1.6~{\rm \upmu m^2}$ FOV sub-region. (c) Magnetic images of the SPIONs' stray fields corresponding to just before switching off the polarizing field ($t=0$) and at several times after ($t=0.12~{\rm s}$, $0.94~{\rm s}$, and $1.94~{\rm s}$). The $t=0$ image is created by averaging together images from the final $0.94~{\rm s}$ before the polarizing field is switched off. The images denoted by $t\rightarrow \infty$ were taken in a separate measurement at a static applied field $B_0=2.0~{\rm mT}$ over several hours, to capture the long-time behavior of the SPIONs at the holding field. (d) Plots of $\Delta m(t)=|\vec{m}(t)-\vec{m}(t\rightarrow\infty)|$. Values of $\vec{m}$ are computed from fits to magnetic images at different values of $t$. Fits to an exponential decay function reveal the Néel relaxation time $\tau_N$ of each SPION. 
    }  
    \label{fig:relax case study}
\end{figure*}

This striking behavior is not accurately captured by ensemble measurements. In~\ref{app:bulkDC}, we compare the $\overline{m}_x(B_0)$ curve of Fig.~\ref{fig:avg_lang}(b) to an ensemble magnetization curve of the same SPIONs taken with a standard SQUID Magnetic Property Measurement System (MPMS), along with ensemble curves reported in the literature of similar SPIONs. The $\overline{m}_x(B_0)$ curve recorded by diamond magnetic microscopy is notably sharper, with a fitted value of $B_c$ that is a factor of 2 to 7 times smaller than the various MPMS measurements. This discrepancy could be associated with differences in sample preparation; for example, substantial interparticle interactions may be present in MPMS ensemble-sample preparations (\ref{app:bulk}).

However, the single-SPION histogram results in Fig.~\ref{fig:avg_lang}(c) are physically reasonable. A standard expression for the characteristic field is $B_c=k_B T/(|\mu \cos{\theta}|)$~\cite{BEA1959}, where $k_B$ is the Boltzmann constant, $T$ is temperature, $\mu$ is the maximum instantaneous SPION magnetic moment, and $\theta$ is the angle between $\vec{B_0}$ and a given SPION's easy axis. Assuming a random easy axis (flat distribution of $\cos{\theta}$), and a narrow distribution for $\mu$ with mean $\overline{\mu}$, the $B_c$ distribution is expected to consist of a sharp peak at ${\sim}k_B T/\overline{\mu}$, followed by a broad $1/B_c^2$ tail, with a median $B_c$ value of ${\sim}2k_B T/\overline{\mu}$. Incorporating the observed median, and assuming room temperature, we infer $\overline{\mu}\approx12~{\rm A{\cdot} nm^2}$, which is consistent with values reported in the literature~\cite{ANT2003,COL2012}. Thus, our interpretation is that SPIONs with sharp, step-like $m_x(B_0)$ curves have easy axes that are nearly aligned with $\vec{B_0}$, while those SPIONs with much more gradual curves have easy axes nearly orthogonal to $\vec{B_0}$. The inverse correlation between $m_x/|\vec{m}|$ and $B_c$ fit parameters at low field support this interpretation (\ref{app:correlate}).

\begin{figure*}[htb]
    \centering
    \includegraphics[width=\textwidth]{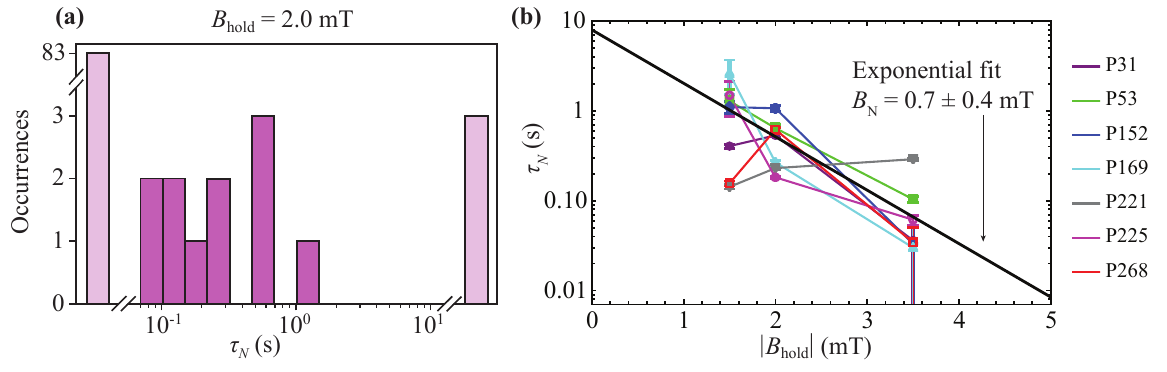} 
    \caption{\textbf{Néel relaxation statistics.} (a) Distribution of Néel relaxation times at $|B_{\rm hold}|=2.0~{\rm mT}$. Eleven SPIONs exhibited relaxation times within the temporal resolution of our setup, ${\sim}0.06\mbox{-}20~{\rm s}$ (dark magenta bars). Relaxation times outside the resolvable range are denoted with light magenta bars. (b) Néel relaxation times of seven SPIONs as a function of $|B_{\rm hold}|$. Error bars for some data points at $|B_{\rm hold}|=3.5~{\rm mT}$ extend below the plot range.
    }
    \label{fig:relax ensemble}
\end{figure*}

\section{Time-resolved Néel relaxation}
\label{sec:relaxation measurement method}

A key enabling feature of our microscope is the ability to image the magnetization \textit{dynamics} of numerous nanoparticles at the same time. This allowed us to characterize the distribution of Néel relaxation properties across the sample.

Thermal fluctuations cause SPION magnetic moments to spontaneously flip, with a characteristic timescale known as the Néel relaxation time, $\tau_N$~\cite{NEE1953,BRO1963}. Here we consider the scenario where a large polarizing field $B_{\rm pol}$ is abruptly switched off, leaving only a much smaller holding field, $B_{\rm hold}$, applied along the same axis. The SPION magnetic moment is initially aligned with $\vec{B}_{\rm pol}$ and decays to a new equilibrium, determined by $B_{\rm hold}$, with a timescale (assuming uniaxial anisotropy) given by:
\begin{equation}
\label{eq:neel}
    \tau_N = \tau_0\exp(\frac{KV}{k_B T}-\frac{B_{\rm hold}}{B_N}).
\end{equation}
In Eq.~\eqref{eq:neel}, $\tau_0$ is a characteristic ``attempt time''~\cite{CHA2021}, $K$ is the anisotropy constant, $V$ is the particle magnetic-core volume, and $B_N$ is an empirical constant that characterizes the field dependence~\cite{NEE1953,BRO1963,COF1995,TAU2012, ADH2024-2}.

 To record the Néel relaxation of individual SPIONs, we performed time-resolved magnetic microscopy using the sequence shown in Figure~\ref{fig:relax case study}(a). A polarizing field, $\vec{B}_{\rm pol} = 31~{\rm mT}\,\hat{x}$ is applied for $\tau_{\rm pol}=2~{\rm s}$ and then abruptly turned off for a time $\tau_{\rm hold}=2~{\rm s}$. The field switching is done via a high-current MOSFET switch, and the field settles with a characteristic $1/e^5$ ring-down time of ${\sim}60~{\rm ms}$ that is limited by the inductance of the coil pair. A second pair of coils is used to apply a constant bias field in the range $\vec{B}_{\rm  hold}=-(1.5\mbox{--}3.5)~{\rm mT}\,\hat{x}$ in the opposite direction as $\vec{B}_{\rm pol}$. Throughout the experiment, magnetic images are continuously recorded, with a repetition time of $\tau_{\rm sweep}=62~{\rm ms}$. The field-switching cycle is done a second time with the microwave frequency sweep directions reversed to suppress rolling-shutter and other timing artifacts (\ref{app:NV}). The overall $8~{\rm s}$ sequence is repeated ${\gtrsim}10^3$ times, and images acquired at the same time $t$, relative to switching $\vec{B}_{\rm pol}$ off, are averaged together to generate high signal-to-noise ratio images.

 Figure \ref{fig:relax case study} shows results for four example particles. SEM images of each particle are shown in Fig.~\ref{fig:relax case study}(b). The corresponding magnetic images ($|B_{\rm hold}|=2~{\rm mT}$) at different times $t$ following the $B_{\rm pol}$ pulse are shown in Fig.~\ref{fig:relax case study}(c).  Each magnetic image is fit (\ref{app:fitting}) to extract the components of $\vec{m}(t)$ for a given particle. The quantity $\Delta m(t)\equiv |\vec{m}(t) - \vec{m}(t\rightarrow\infty)|$ is used to isolate the change in magnetic moment magnitude, $\Delta m(t)$, relative to the (often substantial) static moment, $\vec{m}(t\rightarrow\infty)$, obtained in the presence of $B_{\rm hold}$ alone, see~\ref{app:relax_curves}. The $\Delta m(t)$ curves for each of the four example particles are shown in Fig.~\ref{fig:relax case study}(d), along with fits to an exponential decay function. As can be seen, these four example particles span a wide range of relaxation behaviors, from nearly instantaneous ($\tau_N\lesssim0.06~{\rm s}$) to nearly infinite ($\tau_N\gtrsim20~{\rm s}$). Such a wide particle-to-particle variation is not necessarily surprising, given the exponential dependence of $\tau_N$ on SPION volume, Eq.~\eqref{eq:neel}, but it does validate the importance of single-particle measurements in characterizing the full sample heterogeneity. 

Within the FOV, we extracted the $\Delta m(t)$ curves of $97$ individual SPIONs and fit exponential decay functions to each (\ref{app:relax_curves}). The fitted values of $\tau_N$ are shown as a histogram in Fig.~\ref{fig:relax ensemble}(a). Of this sample, $83$ SPION exhibited relaxation on a timescale too short for us to measure ($\tau_N\lesssim0.06~{\rm s}$), and $3$ exhibited relaxation on a timescale too long for us to measure ($\tau_N\gtrsim20~{\rm s}$). The remaining $11$ SPIONs show a roughly flat distribution. We also performed AC susceptometry on SPIONs from the same batch using ensemble MPMS measurements at variable temperature (\ref{app:bulkAC}). From those measurements, we infer that $\tau_N$ of these SPIONs follows a roughly log-normal distribution with a FWHM of ${\sim}6$ orders of magnitude. Since our diamond magnetic microscopy measurements span only a fraction of the expected FWHM (${\sim}2.5$ orders of magnitude), the observed roughly-flat distribution is in agreement. Interestingly, our single-SPION distribution is most consistent with the MPMS AC susceptibility curve obtained at ${\sim}400~{\rm K}$ (\ref{app:bulkAC}). While the diamond sensor is not this hot (we observe $D\approx2869~{\rm MHz}$, consistent with a temperature of ${\sim}310~{\rm K}$~\cite{ACO2010}), it could be that the SPIONs are locally heated due to the $532~{\rm nm}$ laser excitation~\cite{SAN2018}. Further work will be needed to confirm, as there may be alternative explanations, such as differences due to MPMS ensemble-sample preparation. 

 We imaged SPION relaxation at three values of $|B_{\rm hold}|$ ($1.5, 2, $ and $3.5~{\rm mT}$). Seven SPIONs exhibited a measurable N{\'e}el relaxation time throughout this field range. Their fitted $\tau_N(B_{\rm hold})$ curves are shown in Fig.~\ref{fig:relax ensemble}(b), along with a fit to Eq.~\eqref{eq:neel}. We find good agreement with the predicted exponential dependence, with a fitted decay constant $B_N=0.7\pm0.4~{\rm mT}$, corresponding to a $1/e$ reduction in $\tau_N$. This shows the utility of having a variable $B_{\rm hold}$ in time-resolved relaxation studies, as the $\tau_N$ distribution can be tuned into a measurable range, similar to how temperature can be used in MPMS AC susceptibility measurements to tune distributions to within the detection bandwidth of the SQUID magnetometer.

\section{Discussion}
Our results showcase the potential of diamond magnetic microscopy as a unique platform for high-throughput characterization of the magnetic properties and time-resolved dynamics of individual nanoparticles. As an early study in single-nanoparticle magnetic characterization, a number of our findings will benefit from further investigation and generalization. Examples include the sharp step-like single-SPION magnetization curves, the somewhat-low measured value of $\overline{m}_{\rm sat}$, and the faster N{\'e}el relaxation than expected from room temperature MPMS measurements. 

We envision a number of improvements that can be made to the diamond magnetic microscope that may accelerate validation and future discoveries. The present magnetic image frame rate is ${\sim}16~{\rm Hz}$, which could be increased by several orders of magnitude using faster microwave modulation and image sensors with high-bandwidth electronics~\cite{WOJ2018,PAR2022,TAN2023}. Combined with a reduction in coil ringdown time, either by reducing inductance or using active compensation methods~\cite{SAV2013,DEH2015}, this could enable relaxation measurements that capture $\gtrsim6$ orders of magnitude in $\tau_N$ ($10^{-5}\mbox{-}10^1~{\rm s}$). The microscope's spatial resolution and the magnetic signal strength of SPIONs can be improved by an order of magnitude each using parallel super-resolution magnetic imaging methods~\cite{MOS2024}, though this would likely be accompanied by a slower magnetic image frame rate and lower nanoparticle characterization throughput. The magnetic sensitivity can be improved using pulsed-magnetometry protocols~\cite{HAR2021,KAZ2021} and optimized NV-doped layers~\cite{KLE2016,HEA2020,KEH2021}. The latter could also provide a more precise vertical distribution of NV centers, which can improve absolute quantitation, particularly if combined with the use of calibration standards~\cite{SIE2012,VRE2015}. 

With advances in reproducible sample-substrate preparation~\cite{WAN2018,SCH2021}, future experiments may probe the correlation between SPION morphology and surface chemistry with magnetic properties, such as the anisotropy tensor~\cite{OYA2015}, by correlating diamond magnetic microscopy measurements with high-resolution transmission electron microscopy. The addition of variable temperature control and higher applied magnetic fields would broaden the scope of samples that can be explored. Finally, the techniques presented here can be immediately applied to the study of a wide range of other nanomagnetic phenomena in condensed-matter and biological samples.

In summary, we performed widefield imaging of the stray magnetic fields produced by isolated ${\sim}30~{\rm nm}$ SPIONs using diamond magnetic microscopy. We studied the field-dependent magnetization and time-resolved relaxation of more than a hundred individual SPIONs in parallel, which revealed rich sample heterogeneity. Incorporating the potential improvements outlined above, this platform may accelerate nanomagnetic materials characterization and discovery.

\begin{acknowledgments}
We gratefully acknowledge advice and support from F.~Benito, N.~Arnold, A.~Brearley, D.~Pete, J.~Watt, H.~Hathaway, N.~Adolphi, A.~Cēbers, J.~Cīmurs, J.~T.~McConville, N.~Arnold, and J.~M.~Higbie. \\
\textbf{Competing interests.} The authors declare no competing financial interests.\\
\textbf{Author contributions.} A.~J., A.~L., and V.~M.~A. conceived the idea. P.~K., A.~L., N.~R., B.~A.~R., and V.~M.~A. designed the experiments. A.~L., I.~F., N.~R., and B.~A.~R., built the main experimental apparatus. N.~R. and B.~A.~R. acquired and analyzed magnetic microscopy and SEM data. N.~R., B.~A.~R., and A.~J.~P. acquired and analyzed MPMS data. M.~D.~A., F.~H., N.~M., M.~P.~L., D.~L.~H., T.~K., M.~S.~Z., and A.~M.~M. advised on SPION sample preparation and correlative electron microscopy. D.~L.~H. and T.~K. advised on SPION magnetic properties and modeling.  Y.~S., J.~T.~D., A.~J., T.~K., I.~F., A.~B., A.~L., and J.~S. assisted in theoretical modeling and data interpretation. P.~K., I.~F., and J.~S. wrote the control software. J.~E.~S. and D.~E. conducted TEM imaging and analysis. B.~A.~R. and N.~R. wrote the initial manuscript draft. V.~M.~A. supervised the project. All authors helped edit the manuscript. \\
\textbf{Funding.} This work was supported by the National Science Foundation (DMR-1809800, CHE-1828744, CHE-1945148), National Institutes of Health (R21EB027405, R41MH115884, DP2GM140921), and Moore Foundation EPI-12968. This work was performed, in part, at the Center for Integrated Nanotechnologies, an Office of Science User Facility operated for the U.S. Department of Energy (DOE) Office of Science. Sandia National Laboratories is a multimission laboratory managed and operated by National Technology {\&} Engineering Solutions of Sandia, LLC, a wholly owned subsidiary of Honeywell International, Inc., for the U.S. DOE’s National Nuclear Security Administration under contract DE-NA-0003525. The views expressed in the article do not necessarily represent the views of the U.S. DOE or the United States Government. I.~F. acknowledges support from the Latvian Quantum Initiative under European Union Recovery and Resilience Facility project 2.3.1.1.i.0/1/22/I/CFLA/001. The research at University of Nebraska-Lincoln was performed in part in the Nebraska Nanoscale Facility: National Nanotechnology Coordinated Infrastructure by the National Science Foundation under award no. ECCS: 2025298, and with support from the Nebraska Research Initiative through the Nebraska Center for Materials and Nanoscience.
\end{acknowledgments}

\clearpage
\appendix
\setcounter{equation}{0}
\setcounter{section}{0}
\makeatletter
\renewcommand{\thetable}{A\arabic{table}}
\renewcommand{\theequation}{A\Roman{section}-\arabic{equation}}
\renewcommand{\thefigure}{A\arabic{figure}}
\renewcommand{\thesection}{Appendix~\Roman{section}}
\makeatother

\section{Experimental setup} \label{app:setup}
Excitation light at $532~{\rm nm}$ is sourced from a diode-pumped solid-state laser (MLL-FN-532-500mW) at approximately $500~{\rm mW}$. It then passes through an oscillating diffuser acting as a laser speckle reducer (Optotune, LSR-4C-L) followed by a collimating lens ($f=25.4~{\rm mm}$). A half-waveplate is used to optimize the contrast of the NV optically-detected magnetic resonance lines. The beam passes through an additional lens ($f=200~{\rm mm}$), reflects off a dichroic mirror ($\lambda_{\rm cutoff} \approx 560~{\rm nm}$), and is incident on an oil-immersion microscope objective (NA=1.25) which effectively collimates the beam to a ${\sim}50~{\rm \upmu m}$ diameter in the diamond NV layer. NV fluorescence is collected through the same microscope objective, transmitted through the dichroic mirror, and focused by an infinity-corrected tube lens (Thorlabs ITL200, $f=200~{\rm mm}$) to image onto a water-cooled scientific CMOS image sensor (Hammamatsu ORCA-FLASH 4.0, C11440-22CU). The total magnification of the imaging system is ${\sim}100$x, and a single camera pixel corresponds to a ${\sim}65\times65~{\rm nm^2}$ region on the diamond. 

The diamond sample is positioned using a Thorlabs MDT693B 3-axis open-loop piezo controller. Continuous imaging and localization of a fiducial marker milled into the diamond surface is used to provide feedback to stabilize the diamond position.

Microwaves are generated by two SRS SG384 frequency generators, whose outputs are combined with a power splitter (Mini-Circuits ZN2PD2-63-S+) and amplified (Mini Circuits ZHL-16W-43-S+) with a gain of ${\sim}45~{\rm dB}$. The microwaves are delivered to the NV centers via copper traces deposited onto a glass coverslip. The diamond face opposite to the NV layer is affixed to the coverslip (on top of a portion of a copper trace) using UV-curing adhesive (Norland Optical Adhesive NOA 88).

Two pairs of copper coils are used to apply magnetic fields ranging from $1.5~{\rm mT}$ to $37.2~{\rm mT}$. One pair of coils is water cooled and used to apply the larger fields used in our experiments. This coil pair is switched with a high-current MOSFET switch made by W6PQL Custom Amateur Radio Equipment to provide the $B_{\rm pol}$ field for relaxation measurements. This coil pair has an inductance of ${\sim}40~{\rm mH}$, and we measured its $1/e^5$ ring-down time to be ${\sim}60~{\rm ms}$. This corresponds to the time it takes for a $B_{\rm pol}\approx30~{\rm mT}$ field to settle to ${\sim}0.2~{\rm mT}$ after being switched off. This is the minimum tolerable time following a field switch for us to begin reliably recording optically-detected magnetic resonance (ODMR) spectra without lineshape distortions.

Only one NV axis (out of four possible axes) is used for magnetic measurements. We used a [110]-cut diamond, where two of the NV axes are in plane (i.e. the plane of the largest diamond faces). We selected one of the these in-plane NV axes for magnetic measurements. The diamond was positioned so that this axis was parallel to the applied magnetic field. The laser beam was linearly polarized. A half waveplate was used to align the polarization axis approximately perpendicular to the selected NV axis to maximize ODMR contrast.

An NI PCIe-6321 data acquisition card is used to control the experiment, including synchronizing the microwave sweeps with camera exposures, sending analog signals to the piezo stages, and sending digital signals to the coil current switch.

\section{Diamond chip} \label{app:diamond}
The single-crystal electronic grade diamond used in this study has dimensions $2\times0.2\times0.1~{\rm mm^3}$, with its largest faces polished normal to the diamond's $[110]$ crystallographic direction. It was implanted with $^{15}$N ions with the energies and doses in Table~\ref{tab:dose}. 

\begin{table}[htb]
\caption{Diamond nitrogen-implantation parameters.}
\begin{tabular}{c | c }
 Energy (keV) & $^{15}{\rm N^+~Dose~(cm^{-2}}$) \\ 
 \hline
 10 & $8 \times 10^{12}$ \\
 \hline
 20 & $1.2 \times 10^{13}$ \\
 \hline 
 35 & $1.8 \times 10^{13}$ \\
 \hline 
 60 & $2.8 \times 10^{13}$ \\
 \hline 
 100 & $4.4 \times 10^{13}$ \\
\end{tabular}
\label{tab:dose}
\end{table}

Figure~\ref{fig:srim} shows the expected vertical distribution following implantation, as calculated using the Stopping Range of Ions in Matter (SRIM) package. To convert nitrogen to NV centers, the implanted diamonds were subsequently annealed in vacuum at $800\degree~{\rm C}$ for 4 hours and then $1100\degree~{\rm C}$ for 2 hours, following the recipe in Refs.~\cite{FES2019, KEH2017}.

\begin{figure}[htb]
    \includegraphics[width=\textwidth]{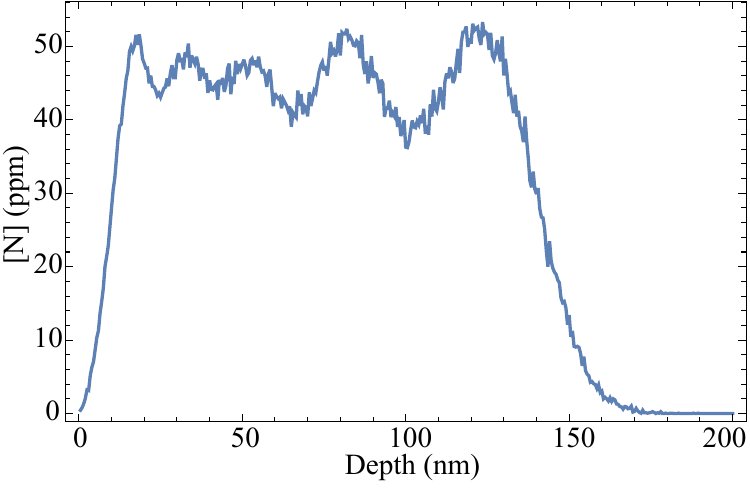}
    \caption{\textbf{Simulated nitrogen depth profile.} The concentration of implanted nitrogen is plotted as a function of depth below the diamond surface. The simulation was performed with the SRIM package.}
    \label{fig:srim}
\end{figure}

\section{SPION properties}
\label{app:SPION}
In contrast to bulk ferromagnetic materials, individual {SPIONs} do not exhibit hysteresis under applied fields that are varied on a timescale much slower than their N{\'e}el relaxation time \cite{PET2010}. In addition, they tend to approach a saturation magnetization at lower applied fields than for typical paramagnets. With sizes on the order of, or smaller, than magnetic domains in bulk iron oxide, {SPIONs} possess a nonzero magnetic moment with a direction determined by their shape, surface chemistry, magnetocrystalline anisotropy, applied magnetic field, and temperature.

\subsection{SPION sample preparation} \label{app:sample}
The {SPIONs} used here were purchased from Ocean Nanotech (SOR30). They were initially suspended in chloroform at a concentration of $25~{\rm mg/mL}$. Prior to depositing on a diamond, the suspension was diluted with hexane by a factor of approximately 4000 and then sonicated for 20 minutes to promote disaggregation. Then, ${\sim}0.3~{\rm \upmu L}$ of the diluted suspension was deposited directly on the diamond surface. The droplet's diameter was larger than the diamond face's $200~{\rm \upmu m}$ dimension and smaller than its $2~{\rm mm}$ dimension. Surface tension kept the droplet from dropping from the pipette tip from which it was dispensed, and so it was touched to the diamond face in order to transfer, and then allowed to dry. In some iterations of this sample preparation, a higher density of {SPIONs} was observed (using the SEM) closer to the edges of the deposited droplet, qualitatively consistent with the so-called coffee-ring phenomenon observed when nanoparticle suspensions are drop-cast \cite{KUM2020}. The FOV containing the {SPIONs} studied here was located ${\sim}100~\upmu$m from these areas of higher {SPION} number density.

\subsection{SEM imaging} \label{app:SEM}
After {SPIONs} are deposited on the diamond, a region near the center of the diamond surface is selected for SEM imaging. {SPIONs} are imaged with an FEI Helios NanoLab 650 SEM to locate single, isolated {SPIONs}. To mitigate charging during SEM imaging, a copper clip mount was used, with the clip pressing on the diamond face itself. The current and accelerating voltage were lowered to $50~{\rm pA}$ and $5~{\rm kV}$ respectively. Even with these measures, charging was found to be highly variable from diamond to diamond. The diamond used for all of the measurements reported in the manuscript (described in \ref{app:diamond}) exhibited sufficiently low charging for imaging, allowing for faithful co-localization with diamond magnetic images after correcting for a small degree of distortion. Images were taken with a magnification of $10000\times$, and each image region of interest (ROI) is approximately $42 \times 28 ~{\rm \upmu m^2}$. 

\begin{figure}[hbt]
    \includegraphics[width=\textwidth]{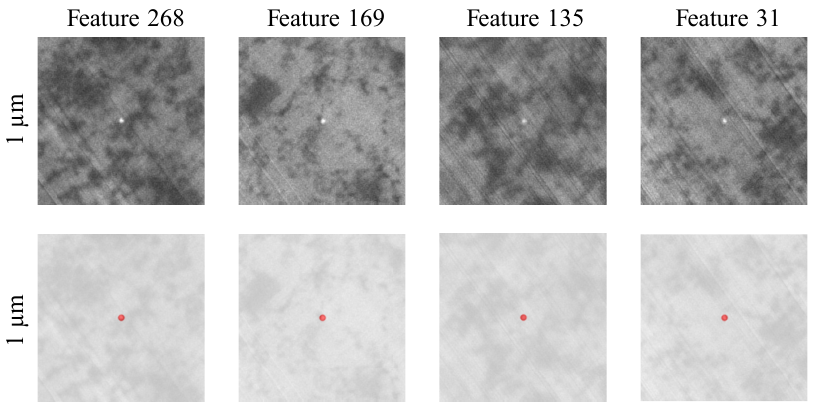}
    \caption{\textbf{SEM image processing.} Raw (top) and processed (bottom) SEM images of the four SPIONs studied in Fig.~\ref{fig:case_study}. Each image is $1 \times 1 ~{\rm \upmu m^2}$.}
    \label{fig:maskSEM}
\end{figure}

For the field of view (FOV) presented in the main text, six of these SEM images were acquired, with some overlap between adjacent images, and stitched together using the software FIJI \cite{SCH2012}, forming a $2\times 3$ grid of images. The ROI of the stitched image was approximately $75 \times 75 ~{\rm \upmu m^2}$. SEM images were taken over a second ROI of similar size elsewhere on the diamond surface, and a similar {SPION} density was observed. The stitched image is then overlaid onto the magnetic microscope image, rotated and resized to match the magnetic image field of view. A small degree of distortion of the SEM image is used to precisely align magnetic dipole patterns with corresponding {SPION} SEM features. During all image transformations, a fiducial marker milled into the diamond surface (visible in both the SEM and magnetic images) is used to ensure that the overlay is accurate to within $\lesssim1~{\rm \upmu m}$. The final SEM image is then cropped to match the magnetic image FOV.

The SEM images have a spatial resolution of $\lesssim10~{\rm nm}$. Together with the substantial image contrast between {SPIONs} and diamond, this allowed {SPIONs} to be identified by eye in the images. In the final stitched SEM image of the FOV presented in the main text, single and aggregate {SPIONs} were marked in color manually to distinguish them from the diamond surface. The intensity of the background (diamond surface) pixels was reduced by 70\% for visual clarity. A comparison of SEM images of several {SPIONs} before and after the image processing is shown in Fig.~\ref{fig:maskSEM}.

\begin{figure}[hbt]
    \includegraphics[width=\textwidth]{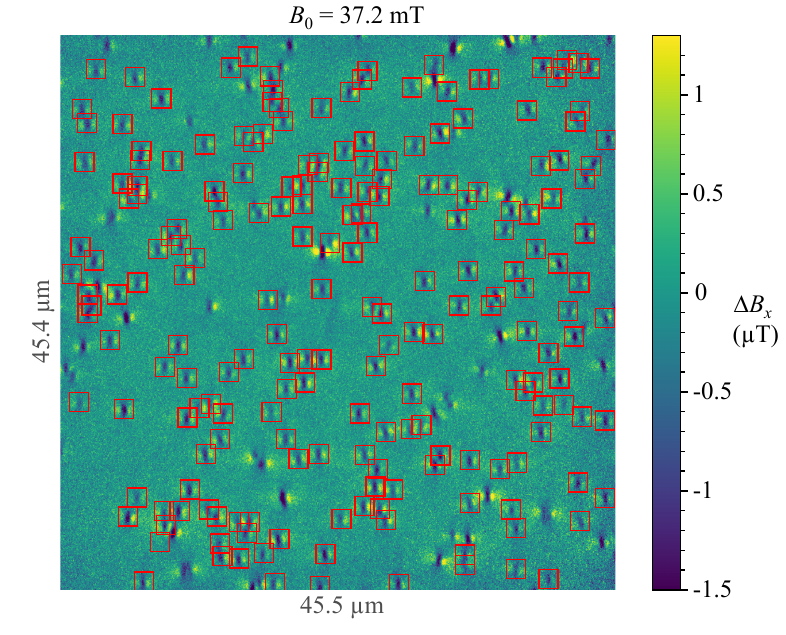}
    \caption{\textbf{Magnetic field map.} Example diamond magnetic microscopy image. Red boxes surround the magnetic features of \textit{single} SPIONs located $\gtrsim 1~{\rm \upmu m}$ from its nearest neighbor, as determined by SEM co-localization. 207 such SPIONs were located in this FOV. Each feature is assigned a unique feature number, but the numbers are not shown on the plot for visual clarity.
    }
    \label{fig:boxed}
\end{figure}

Following the initial processing, we used Wolfram Mathematica's morphological image analysis function, ``ComponentMeasurements'', to locate the pixel coordinates of the centroid of each single {SPION} or {SPION} aggregate. A list of these image coordinates was generated, together with a corresponding list identifying each feature as a single {SPION}, {SPION} aggregate, or unidentified {non-SPION} particulate. Additionally, it was noted whether each feature was ${\gtrsim}1.5~{\rm \upmu m}$ or ${\gtrsim}1~{\rm \upmu m}$ from its nearest neighboring feature. (Distances within the SEM image were quantified by measuring the number of pixels corresponding to the scale bar generated by the SEM. Some error is introduced by the rotation and distortions described above, as well as sample charging during image acquisition, but the error is small enough that feature co-localization with magnetic images was still possible.) 

Figure~\ref{fig:boxed} shows a magnetic image taken with an applied field of $B_0 = 37.2~{\rm mT}$. The image is annotated with boxes around the magnetic features corresponding to single {SPIONs} that are each $\gtrsim 1~{\rm \upmu m}$ from their nearest neighbor as determined by SEM imaging.

We imaged more than three ${\sim}75\times 75~{\rm \upmu m^2}$ ROIs with the SEM, taken over several repetitions of the same SPION drop-casting process. Some of these regions were not studied with the magnetic microscope due to a low number of isolated single SPIONs (owing to a SPION-suspension number density that was either too high or too low). For other regions, magnetic images were taken over a range of applied fields, but the full set of studies described in the main text were unable to be completed due to degradation of the optical adhesive under the diamond after exposure to the $532~{\rm nm}$ laser light for ${\gtrsim}4$ weeks.

\begin{figure}[hbt]
    \includegraphics[width=\textwidth]{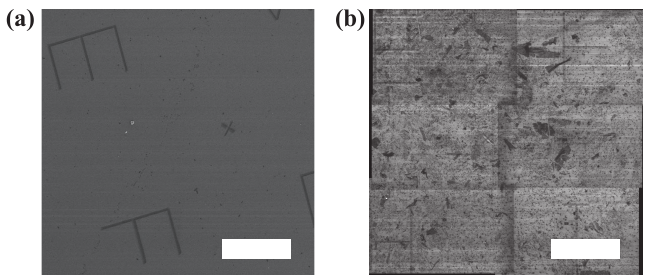}
    \caption{\textbf{Diamond surface before and after drop-casting.}
    (a) SEM image of the diamond surface after cleaning but before drop-casting SPIONs. (b) SEM image of the diamond surface after SPION deposition. The scale bars in both (a) and (b) are $20~{\rm \upmu m}$.
    }
    \label{fig:hexane}
\end{figure}

In SEM images with SPIONs on diamond, for example Fig.~\ref{fig:maskSEM}, there are pronounced contrast ``blemishes'' of varying size. To investigate their origin, we acquired SEM images of the clean diamond surface prior to depositing SPIONs on the diamond surface. The diamond was cleaned in a solution of 1:1:1 nitric:sulfuric:perchloric acid at $\sim 200 \degree$ C for several hours followed by rinsing with deionized water and then isopropyl alcohol. Figure \ref{fig:hexane}(a) shows an SEM image of the diamond, taken after this cleaning procedure. Next we drop-cast SPIONs on to the diamond surface and acquired a second image of a similar region of the diamond, Fig.~\ref{fig:hexane}(b). Comparison of these images suggests that most of the dark blemishes are caused by residues from the hexane in the initial SPION suspension. Some features may also be due to inadvertent contact between the diamond and its container.

\subsection{TEM imaging}

\begin{figure*}[hbt]
    \includegraphics[width=\textwidth]{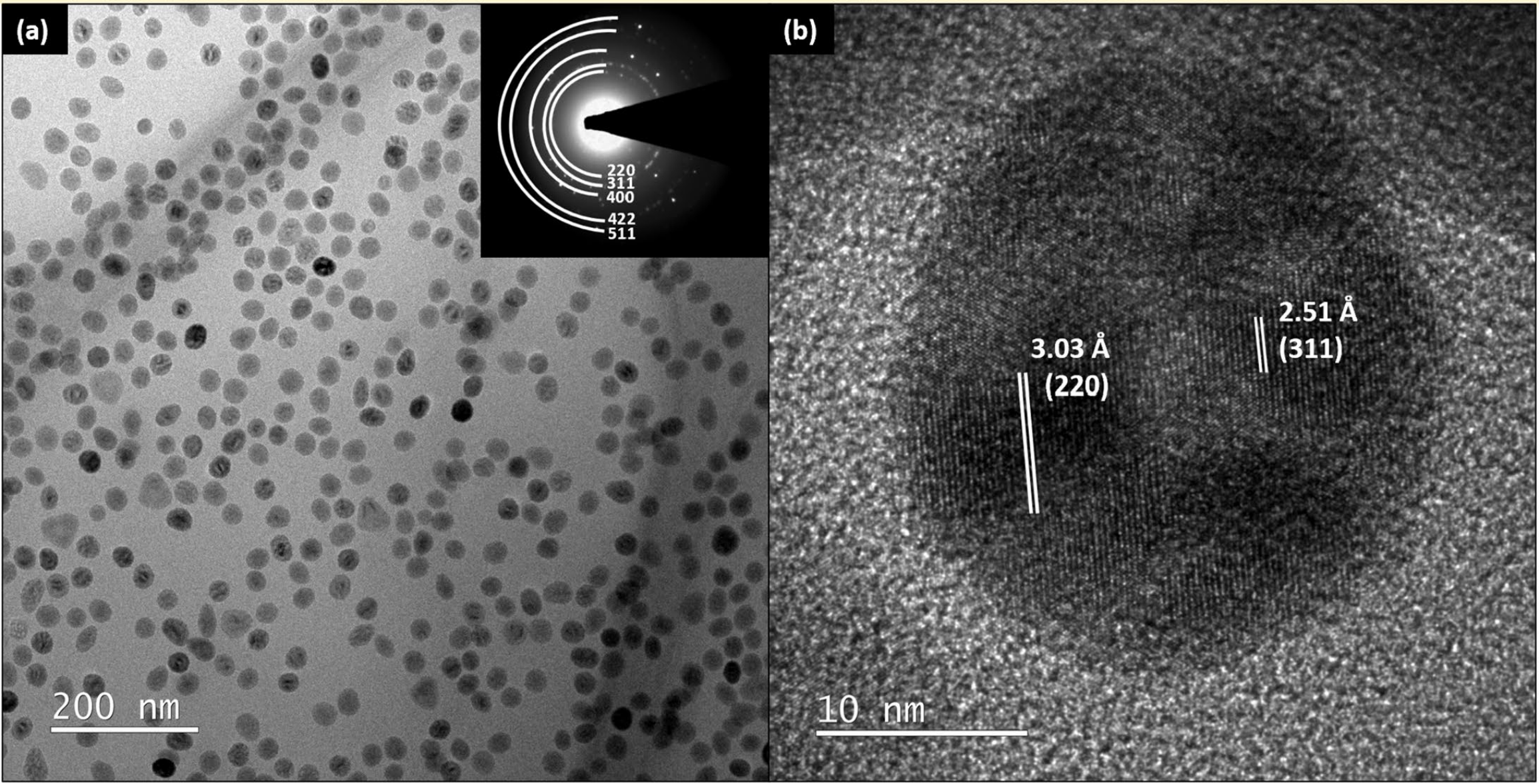}
    \caption{\textbf{Transmission electron microscopy (TEM) images of SPIONs.}
    (a) Low-magnification TEM image of SPIONs used to determine shape and size distribution. Inset: selected area electron diffraction pattern consistent with face-centered cubic Fe$_3$O$_4$. (b) High-resolution transmission electron microscopy image of a SPION showing indexed lattice fringes. The relative orientation of the planes is inconsistent with the SPION being a single crystal.
    }
    \label{fig:tem}
\end{figure*}

\label{app:tem}

The standard drop-casting method, as described in \cite{ZHE2014}, was used to prepare the SPIONs for transmission electron microscopy (TEM) analysis. This process involved extracting a small volume of the SPIONs dispersed in chloroform using a micropipette and depositing it onto a TEM copper grid. The sample was then allowed to air dry for 3 hours to ensure the complete evaporation of the solvent before TEM analysis. TEM analysis was conducted using the FEI Tecnai Osiris operating at 200 kV. As seen in Fig.~\ref{fig:tem}(a), the SPIONs exhibit a spherical morphology with a nearly uniform particle diameter. Using an image segmentation program, we determined the mean particle diameter to be $30~{\rm nm}$, with a standard deviation of ${\sim}4~{\rm nm}$. A selected area electron diffraction pattern, inset of Fig.~\ref{fig:tem}(a), reveals distinct diffraction rings indexing to face-centered cubic Fe$_3$O$_4$. High-resolution images, Fig.~\ref{fig:tem}(b), show lattice spacings consistent with the $\{220\}$ and $\{311\}$ interplanar spacings for face-centered cubic Fe$_3$O$_4$ crystal structure, corroborating the diffraction results. The high-resolution TEM images also reveal evidence that the SPIONS were polycrystalline. Lattice fringes from a given plane do not extend across the diameter of the nanoparticle, and the observed $(220)$ and $(311)$ planes in Fig.~\ref{fig:tem}(b) are nearly parallel, which structurally cannot occur in a single crystal.

\section{Dual-resonance diamond magnetic microscopy} \label{app:NV}

\subsection{Concept}
\label{app:dualRes}

Dual-resonance continuous-wave ODMR imaging is performed by simultaneously sweeping two microwave tones about the NV ground-state transition frequencies. One tone, $f_1(t)$, is swept about the $m_s=0 \leftrightarrow-1$ transition frequency (in the absence of a magnetic sample), denoted $f_-$, while the other tone $f_2(t)$ is simultaneously swept about the $m_s=0\leftrightarrow +1$ transition, $f_+$.
One of these sweeps is done in the direction of increasing frequency, while the other is swept with decreasing frequency. A sweep deviation $\delta=12~{\rm MHz}$ is chosen such that each sweep spans the frequency range $f_{\pm} - \delta$ to $f_{\pm} + \delta$. Together with continuous optical excitation with 532 nm laser light, this results in a single dip in fluorescence brightness which reports on the local magnetic field. 

At any time $t$ during the microwave frequency sweep, a detuning $\Delta$ is defined as:
\begin{equation}
f_1(t)-f_-= f_+ - f_2(t) \equiv \Delta,
\end{equation}
where it is assumed that the center of each microwave tone's sweep is tuned to its respective $f_{\pm}$ resonance. Since the microwaves are swept at a constant rate, the fluorescence-intensity as a function of time, for a given pixel, can be converted to an ODMR spectrum (fluorescence-intensity as a function of $\Delta$) with a linear mapping. In the absence of additional fields, the ODMR peak has a central frequency at $\Delta\approx0$. 

Since the frequency sweeps are performed in opposite directions, a change in temperature does not change the resonance central frequency. Temperature shifts do broaden the dual-resonance ODMR line somewhat, but for shifts of $\lesssim10~{\rm K}$, the degree of broadening is much smaller than the ODMR linewidth and has a negligible effect on the spectrum. 

On the other hand, a shift in the magnetic field along the NV axis due to the sample, $B_x$, results in a shift of the resonance central frequency given by:
\begin{equation}
\label{eq:fc}
    \Delta_c = \pm\, \gamma_{\rm nv} B_x,
\end{equation}
where the sign on the right-hand side of Eq.~\eqref{eq:fc} depends on the directions of the frequency sweeps. For example, if $f_1(t)$ is swept down in frequency and $f_2(t)$ is swept up, the sign is positive. In order to measure $B_x$, we fit a given pixel's ODMR spectrum to a Lorentzian function of the form:
\begin{equation}
    F(\Delta)=F_0(1-C\frac{(\Gamma/2)^2}{(\Delta - \Delta_c)^2 + (\Gamma/2)^2}),
\end{equation}
where $C$ is the ODMR contrast, $\Gamma$ is the full-width at half-maximum (FWHM) linewidth, $F_0$ is the off-resonance fluorescence intensity, and $\Delta_c$ is the resonance's central frequency with respect to the scan's center. For a single pixel, the typical count rate was $F_0\approx40~{\rm kilocounts}$ for a $5.2~{\rm ms}$ exposure time, the ODMR contrast was $C\approx0.03$, and the FWHM linewidth was $\Gamma\approx12~{\rm MHz}$.

\subsection{Sensitivity enhancement}
\label{app:dualsens}

In the main text, we report a ${\sim}1.4$-fold improvement in sensitivity when using dual-resonance magnetometry, as compared to the typical single-resonance approach. This improvement is based on the observed change in the ODMR spectrum upon driving both resonances simultaneously. Here, we discuss factors contributing to the observed increase in ODMR contrast. 

The minimum detectable magnetic field is directly proportional to the ratio $\Gamma/C$, where $\Gamma$ is the ODMR FWHM linewidth and $C$ is the ODMR contrast~\cite{ACO2010APL}. In our experiments, we first minimized this ratio (by adjusting microwave power) under single-resonance driving. Next, we did the same optimization under dual-resonance driving. We observed that the optimized $\Gamma/C$ ratio decreased by a factor of ${\sim}1.4$ when we applied dual resonance (compared to single resonance), indicating a ${\sim}1.4$-fold improvement in sensitivity. Specifically, we observed that $C$ increased by a factor of ${\sim}1.4$ while $\Gamma$ hardly changed. This improvement in sensitivity turned out to be quite helpful in reducing the experimental integration times.

In an ideal Ramsey-type pulsed ODMR measurement, there should be no improvement in fluorescence contrast when using dual-resonance excitation~\cite{FAN2013}. The situation in continuous-wave ODMR is more complicated, as there is a rich interplay between the optical excitation rate ($I_0$), microwave Rabi frequency ($\Omega_R$), NV spin dephasing rate ($1/T_2^{\ast}$), and NV longitudinal spin relaxation rate ($1/T_1$)~\cite{DRE2011,JEN2013}. In the limit $1/T_2^{\ast} \gg \Omega_R \gg I_0 \gg 1/T_1$, we can apply a classical rate equation model to gain intuition about the dynamics. In this case, as shown in~\cite{FES2020}, under dual resonance conditions, the population in $m_s=0$ approaches $1/3$. On the other hand, in the single resonance case, the population in $m_s=0$ would be ${\sim}1/2$. Thus, in this toy model, the contrast improves by a factor of ${\sim}4/3$. Another extreme is the situation where the NV dephasing rate is low ($T_2^{\ast}$ is long) and $\Omega_R$ is relatively high, as explored in Ref.~\cite{SHI2022}.  There, it was observed that the contrast under dual resonance excitation can actually decrease (compared to single resonance), which was attributed to complicated dynamics that counteract population trapping in coherent dark states.

Experiments have reported various other contrast enhancements: a factor of ${\sim}1.3$ in Ref.~\cite{FES2020} and a factor of 1.5-1.6 in Ref.~\cite{YU2024}, though we cannot confirm if the microwave field amplitude was optimized in every case. The exact solution for arbitrary $T_2^{\ast}$, $\Omega_R$, $I_0$, and $T_1$ would require solving density-matrix quantum master equations~\cite{DRE2011,JEN2013}, which could make for an interesting future study.

\subsection{Sequences and timing}
\label{app:timing}
Each microwave frequency sweep consists of 12 frequencies spaced over a $24~{\rm MHz}$ full span. Twelve fluorescence images are collected, one at each detuning (see \ref{app:dualRes}). The time allocated for each frequency step is equal to the exposure time $\tau_{\rm exp}=5.2~{\rm ms}$ for all measurements in this paper.

Each image taken with $f_2(t)$ swept up in frequency (while $f_1(t)$ is swept down in frequency) is subtracted from an image in which $f_2(t)$ is swept down while $f_1(t)$ is swept up. This allows for removal of magnetic image artifacts resulting from the interplay between the camera's rolling shutter and the changing microwave frequencies.

\begin{figure*}[htb]
    \centering
    \includegraphics[width=0.9\textwidth]{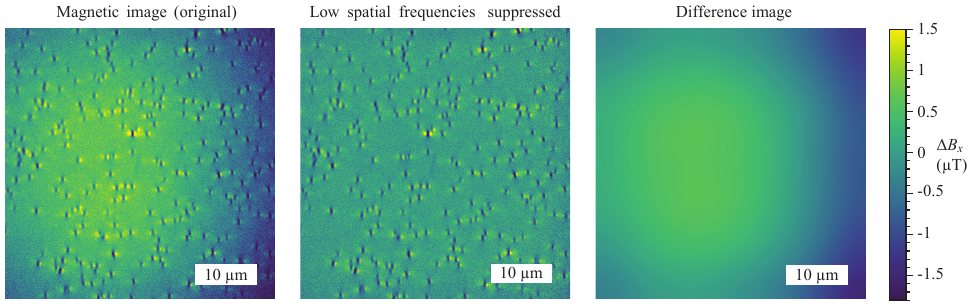}
    \caption{\textbf{Magnetic image spatial filtering.} (left) A raw magnetic image exhibits low-frequency artifacts with spatial periods on the order of the FOV dimension (${\sim}45~{\rm \upmu m}$). (middle) Filtered magnetic image that suppresses the unwanted low-frequency background. The raw image (left) is convolved with a Gaussian kernel of size $401\times 401$ pixels and standard deviation 50 pixels to obtain a blurred image (right). The filtered magnetic image is obtained by subtracting the blurred image from the raw magnetic image, effectively passing only higher spatial frequencies. 
    }
    \label{fig:imageProc}
\end{figure*}

For measurements taken at static applied field, the sweep directions are alternated every ${\sim}4~{\rm s}$ and the entire cycle is repeated for a time on the order of one hour. Within a set of microwave sweeps of the same directions, all fluorescence images corresponding to a particular microwave detuning are averaged together. This results in a pair of magnetic images, one for each sweep direction, which are then subtracted.

For time-resolved imaging, one set of coils provides a pulsed applied field $B_{\rm pol}$ that is on for $2~{\rm s}$ and off for $2~{\rm s}$ for a total cycle period of $4~{\rm s}$. During a given cycle period, 33 magnetic images are taken during the ``field on" duration and 33 images are taken during the ``field off" duration. The holding field $B_{\rm hold}$ in the second set of coils (directed opposite to $B_{\rm pol}$) is kept on at all times. For the relaxation analysis, the first two magnetic images during each ``field off" duration are dropped to ensure that the field is completely settled following the ${\sim}60~{\rm ms}$ ringdown in the pulsed coils. The microwave sweep directions are alternated between each on-off cycle of the field. For a given time point, $t$, and a given set of microwave sweep directions, all fluorescence images corresponding to a particular microwave detuning are averaged together. For each time point, the resulting pair of magnetic images is then subtracted to mitigate rolling-shutter artifacts as described above.

\subsection{Parallel ODMR fitting}
\label{app:fitting}
Each magnetic image is obtained by fitting the ODMR spectrum of each pixel in the FOV ($699\times700$ pixels) for each of the two sweep directions, corresponding to a total of 978600 Lorentzian fits. The fitting procedure was performed using a graphics card (Asus Nvidia GeForce RTX 3060; 3584 CUDA cores; 12 GB GDDR6) in order to parallelize the computation. The fits were computed with CUDA, via the Python package Gpufit \cite{PRZ2017}. This allowed the 978600 fits for a single magnetic image to be completed in approximately 10 seconds.

A number of ${\sim}45 \times 45 ~{\rm \upmu m}^2$ FOVs were imaged in this manner. The complete analysis of SPIONs presented in the main text was completed only for a single FOV. For other FOVs, only a partial analysis was possible due to laser damage to the UV-curing adhesive layer (see \ref{app:setup}) after weeks of exposure.

\section{Image processing} \label{app:processing}

The widefield magnetic images obtained as described in \ref{app:NV} exhibit magnetic gradients across the FOV. These low-spatial-frequency magnetic patterns vary from image to image. We speculate that they may arise from real magnetic field gradients, or they may be artifacts due to inhomogeneity of the microwave or optical fields or interplay between the camera rolling shutter and the MW frequency sweeps.  To suppress these patterns, we apply a modest high-pass filter. Specifically, a Gaussian blur is applied to the image using a 401$\times$401 pixel kernel with a standard deviation of 50 pixels. The blurred image is subtracted from the original magnetic image. An example of the filtering procedure is shown in Figure \ref{fig:imageProc}.

In the time-resolved magnetic images, a slight blur is applied to the filtered image to suppress high-spatial-frequency noise (corresponding to length scales well below the diffraction limit). This Gaussian blur uses a kernel area of $13\times13~{\rm pixels^2}$ and a standard deviation of 1 pixel. These values were chosen by examining magnetically bright features (corresponding to clusters of SPIONs) to ensure that the blur does not significantly alter the fits of magnetic moment. Specifically, magnetic images of these features were convolved with Gaussian kernels of varying kernel sizes and standard deviations, and line cuts across each blurred image were compared with a line cut across the unblurred image. The values of $13\times13~{\rm pixels^2}$ kernel area and 1 pixel standard deviation did not appreciably increase the width or peak-to-peak amplitude of these magnetic features.

\section{SPION stray field fits} \label{app:dipoleFits}
As described in Sec.~\ref{app:SEM}, magnetic features of single SPIONS were co-localized with SEM images. In the magnetic images, sub-regions of interest (${\sim}1.5\times1.5~{\rm \upmu m^2}$) were cropped about each isolated SPION feature. For each sub-region, a two-dimensional fit was performed to extract the magnetic moment vector $\vec{m}$ of the particle. The fit function is given by:
\begin{equation}
\label{eq:convolve}
    B_x(x,y) = \frac{1}{5}\sum_{i=1}^5 B_{\textrm{dip},x}(x,y,z_i) \ast \textrm{PSF}(x,y),
\end{equation}
where
\begin{equation}
\begin{aligned}
\label{eq:dipole field}
    B_{\rm dip,x}(x,y,z) =\,&\frac{3 (x{-}x_0)}{r^5}[m_x(x{-}x_0){+}m_y(y{-}y_0){+}m_z z]\\ &-m_x/r^3 + B_{\rm{off}}.
\end{aligned}
\vspace{2 mm}
\end{equation}
The summand in Eq.~\eqref{eq:convolve} is a convolution of the $x$-component of the SPION dipole field, $B_{\rm dip,x}(x,y)$, with the microscope optical point-spread function, ${\rm PSF}(x,y)$. This expression is averaged over 5 different $z$ depths, ($z_i=-\{60,90,120,150,180\}~{\rm nm}$) selected to represent the relevant NV layer distribution, as described in~\ref{app:diamond}. The $60~{\rm nm}$ standoff is chosen to account for the SPION radius (${\sim}15~{\rm nm}$), a thin hydrocarbon layer between SPION and diamond (${\sim}10~{\rm nm}$), a (${\sim}15~{\rm nm}$) depleted layer near the diamond surface where implanted nitrogen is not efficiently converted to negatively-charged NV centers~\cite{MOS2024}, and an additional (${\sim}20~{\rm nm}$) ``dead layer'' where the NV ODMR signal is not expected to contribute. The ``dead layer'' comes from the expectation that NV centers closer than ${\sim}60~{\rm nm}$ to a SPION center exhibit large enough ODMR line shifts that their contribution to the voxel-averaged ODMR lineshape is inadequately captured by a least-squares Lorentzian fit. Thus, while the NV layer is predicted to have a thickness of ${\sim}130~{\rm nm}$ (\ref{app:diamond}), we sample the distribution in five planes extending over ${\sim}120~{\rm nm}$. This difference in NV distribution length has a relatively minor impact on the simulated magnetic field profiles, whereas other sources of uncertainty (for example, deviations from the expected SRIM profile) could play a larger role.

The function ${\rm PSF}(x,y)$ in Eq.~\eqref{eq:convolve} is a Gaussian function that accounts for the diffraction-limited point-spread function of the microscope as well as any additional blurring caused by mechanical motion during magnetic image acquisition. For a given magnetic image, ${\rm PSF}$ is found empirically by analyzing the pattern width of magnetically bright features (corresponding to small clumps of SPIONs) and is held fixed when analyzing features of individual SPIONs. In Eq.~\eqref{eq:dipole field}, the fit parameters are the components of the magnetic moment ($m_x,\, m_y,\, m_z$), the coordinates of the dipole ($x_0,\, y_0$), and a constant offset term, $B_{\rm off}$. Here $r=\sqrt{(x-x_0)^2+(y-y_0)^2+z^2}$ is the displacement magnitude.

\section{Correlating single-SPION magnetic properties}
\label{app:correlate}
Our approach allows us to track individual SPIONs throughout a large suite of magnetic measurements. We can thus search for correlations in the extracted magnetic properties. Figure~\ref{fig:corr-plots}(a) plots $B_c$ vs. $m_{\rm sat}$ for 96 of the SPIONs in the field of view. No strong correlations are observed. Figure~\ref{fig:corr-plots}(b) plots $B_c$ vs. $\tau_N$. Once again, no clear correlation is observed. 

\begin{figure}[htb]
    \includegraphics[width=\textwidth]{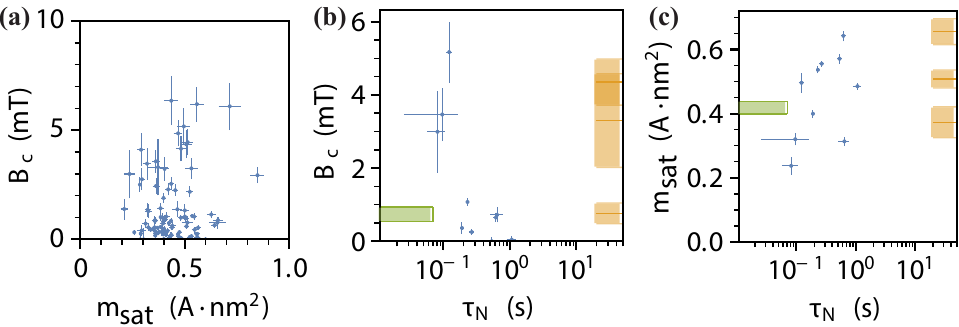}
    \caption{\textbf{Correlation plots.} (a) Scatter plot of the Langevin fit parameters $B_c$ versus $m_{\rm sat}$ for 96 of the SPIONs in the sample. The error bars denote fit uncertainties. (b) Plot of $B_c$ (measured by field-dependent magnetization) versus Néel relaxation time $\tau_N$ (measured by time-resolved magnetic relaxation). The horizontal orange lines and rectangles show the $B_c$ values and fit uncertainty for SPIONs with $\tau_N \gtrsim20~{\rm s}$. The green rectangle is centered about the median $B_c$ for SPIONs with $\tau_N \lesssim0.06~{\rm s}$. Its upper and lower edges are the 60th and 40th percentiles of the $B_c$ values for this group of SPIONs. (c) $m_{\rm sat}$ versus $\tau_N$. The orange and green rectangles are defined in the same manner as in part (c) but for $m_{\rm sat}$.}
    \label{fig:corr-plots}
\end{figure}

Figure~\ref{fig:corr-plots}(c) plots $m_{\rm sat}$ vs. $\tau_N$. As $m_{\rm sat}$ is expected to be proportional to SPION volume, and $\tau_N$ has an exponential dependence on SPION volume, we anticipate the two to be positively correlated. There is a considerable scatter in the points in Fig.~\ref{fig:corr-plots}(c), and only a relatively small number of SPIONs have a well-resolved value of $\tau_N$. Nonetheless, the data are not inconsistent with the expected correlation.

\begin{figure}[htb]    
\includegraphics[width=0.9\textwidth]{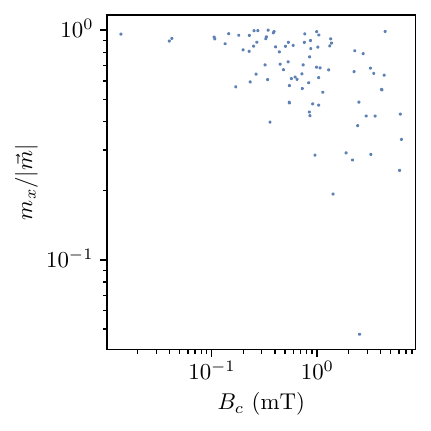}
    \caption{\textbf{Easy axis angle.} An approximate value of $\cos{\theta}$ is given by $m_x/|\vec{m}|$, where $\theta$ is the angle of the easy axis with respect to the applied field direction and $\vec{m}$ is extracted from fits to SPIONs at the lowest practical applied field ($B_0=1.5~{\rm mT}$). A scatter plot of $m_x/|\vec{m}|$ vs. $B_c$ is shown for ${\sim}95$ SPIONs. As the standard expression for characteristic field is $B_c=k_B T/(\mu\cos{\theta})$, the approximately inverse relationship observed in this plot is expected.
    }
    \label{fig:mpar-Bc}
\end{figure}

As discussed in the main text, the characteristic field for a SPION can be written $B_c = k_B T/\left(\mu \cos{\theta} \right)$, where $\theta$ is the angle of the easy axis relative to the applied magnetic field, $T$ is the temperature, and $\mu$ is the instantaneous SPION magnetic moment. For applied fields of sufficiently small magnitude, $\cos{\theta}$ is approximately given by $m_x/|\vec{m}|$, where the applied field is in the $x$-direction. Figure \ref{fig:mpar-Bc} plots $m_x/|\vec{m}|$ at an applied field of 1.5 mT (the lowest applied field used in our measurements) against $B_c$ for $\sim 95$ SPIONs. The data are consistent with an inverse dependence on $B_c$, as anticipated.

\section{Comparison with bulk measurements} \label{app:bulk}

\subsection{Bulk sample preparation and measurement}
\label{app:MPMSintro}
A Quantum Design MPMS-3 SQUID magnetometer was used to characterize SPIONs of the same batch as those imaged by diamond magnetic microscopy. Samples were prepared by suspending SPIONs in hexane, sonicating for approximately $10$ minutes to discourage particle aggregation, and depositing the suspension on a cotton swab. The cotton swab was fixed in a plastic straw sample holder for measurement inside the MPMS. An alternative sample preparation we used for bulk SPION characterization involved freeze-drying a SPION suspension in mannitol \cite{DIE2016}. 

\subsection{Field-dependent magnetization curves}
\label{app:bulkDC}

\begin{figure}[htb]
    \includegraphics[width=\textwidth]{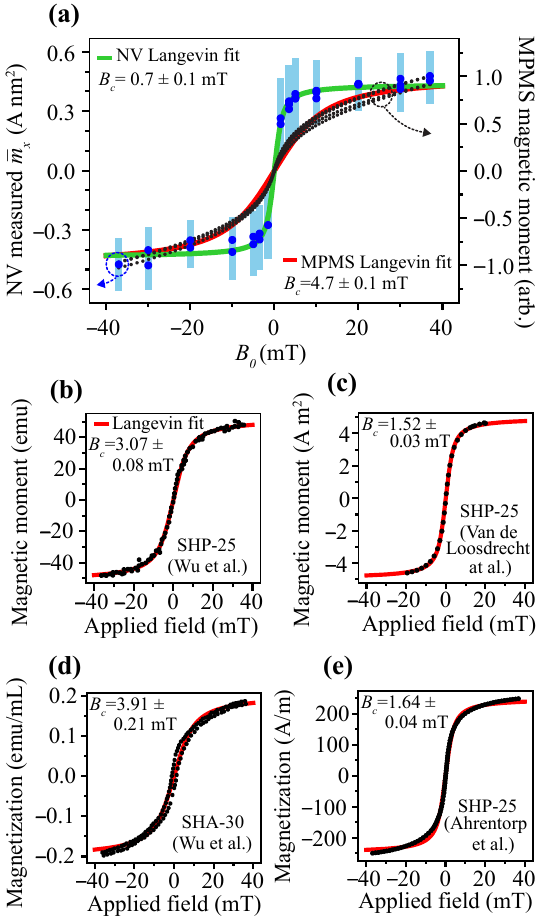} 
    \caption{\textbf{Magnetization measurement comparison.} (a) SPION magnetic moments versus applied magnetic field as measured with NV-magnetic microscopy, averaged over the isolated single SPIONs in the FOV (blue points and green fit). The values shown are the component of the magnetic moment along the direction of the applied magnetic field. SPION magnetic moments versus applied magnetic field, as measured with the Magnetic Property Measurement System (MPMS) SQUID magnetometer (black points and red fit). (b,c) SPION magnetic moments versus applied magnetic field, as measured in \cite{WU2015, VAN2019}. (d,e) SPION magnetization versus applied magnetic field, as measured in \cite{WU2021, AHR2015}.}
    \label{fig:lit_hysteresis}
\end{figure}

 The MPMS was used to measure the magnetic flux from a bulk sample of SPIONs along the direction of the applied field. The applied field is swept up and down from $\pm37~{\rm mT}$ several times, mimicking the procedure used by diamond magnetic microscopy. Figure~\ref{fig:lit_hysteresis}(a) shows a plot of a typical magnetization curve obtained at $T=400~{\rm K}$. These data used a SPION sample absorbed and dried on a cotton swab, prepared with a suspension density of 0.25 mg/mL. Hints of hysteresis are observed, but the data are still reasonably described by a Langevin function, Eq.~\eqref{eq:langevin}. The fit reveals a fairly large characteristic field, $B_c=4.7\pm0.1~{\rm mT}$, where the uncertainty is the fit standard error. This is a factor of ${\sim}7$ larger than that observed by diamond magnetic microscopy ($B_c=0.7\pm0.1~{\rm mT}$), despite using SPIONs from the same initial suspension.

 We repeated the bulk MPMS measurements under a variety of conditions: i) a dilute suspension of the same density (0.25 mg/mL), drop-cast on a cotton swab, measured at three different temperatures (300 K, 360 K, and 400 K); ii) a 10-fold higher suspension density (2.5 mg/mL), drop-cast on a cotton swab, measured at 300 K; iii) A 100-fold higher suspension density (25 mg/mL), drop-cast on a cotton swab, measured at 300 K; and iv) the 25 mg/mL suspension immobilized in freeze-dried mannitol, measured at 300 K. In all cases, a similarly small amount of hysteresis was observed, and the fitted characteristic field lies in the range of approximately $4.7\mbox{-}8~{\rm mT}$. As discussed in the main text, we are unsure of the source of this discrepancy but speculate it could be due to differences in sample preparation between MPMS bulk measurements and diamond magnetic microscopy.
 
Figure~\ref{fig:lit_hysteresis}(b-e) shows four different SPION-ensemble magnetization curves reported in the literature~\cite{WU2015, VAN2019, AHR2015, WU2021}, where the SPIONs were of a similar size and composition to those studied here. Our MPMS measurement of $B_c = 4.7 \pm 0.1~{\rm mT}$ at $400~{\rm  K}$ lies just outside the range of $1.52\mbox{-}3.91~{\rm mT}$ reported in these papers. The variation could be due to differences in SPION size or composition or potentially contributions from inter-SPION interactions, owing to differing sample preparations.

The mean of the single-SPION $m_{\rm sat}$ distribution [$\overline{m}_{\rm sat}= 0.44~{\rm A{\cdot}nm^2}$, see Fig.~\ref{fig:avg_lang}(c)] can also be compared with literature values. Wu et al.~\cite{WU2021} reported a saturation magnetic moment of $\overline{m}_{\rm sat}=2.5~{\rm A{\cdot}nm^2}$ per nanoparticle for 30-nm-diameter Ocean Nanotech SHA-30 SPIONs, and $\overline{m}_{\rm sat}=1.6~{\rm A{\cdot}nm^2}$ per nanoparticle for 25-nm-diameter Ocean Nanotech SHA-25 SPIONs. Ahrentorp et al.~\cite{AHR2015} reported $\overline{m}_{\rm sat}=2\mbox{-}3.4~{\rm A{\cdot}nm^2}$ for 25-nm SPIONs suspended in liquid. Mosavian et al.~\cite{MOS2024} used a super-resolution diamond magnetic microscopy technique to record $m_x=0.86~{\rm A{\cdot}nm^2}$ for a single Ocean Nanotech SOR30 SPION at an applied field of $30~{\rm mT}\,\hat{x}$. Our diamond magnetic microscopy measurements on 30-nm SPIONs yield $\overline{m}_{\rm sat}= 0.44~{\rm A{\cdot}nm^2}$, which is somewhat smaller than these values.

There are several possible reasons why our method might record spuriously lower measured magnetic moments. Diamond magnetic microscopy can be affected by pixel cross-talk in which signals from distant NV centers contribute to the ODMR spectrum of a camera pixel near a SPION \cite{MOS2024, SCH2022}. Additionally, it is possible that the distribution of NV depths in the diamond is broader and deeper than estimated by SRIM; for example, due to ion channeling effects. This could be prevented by using delta-doped NV layers \cite{KLE2016, HEA2020,KEH2021}. 

\subsection{AC susceptibility}
\label{app:bulkAC}

\begin{figure}[htb]
    \includegraphics[width=0.9\textwidth]{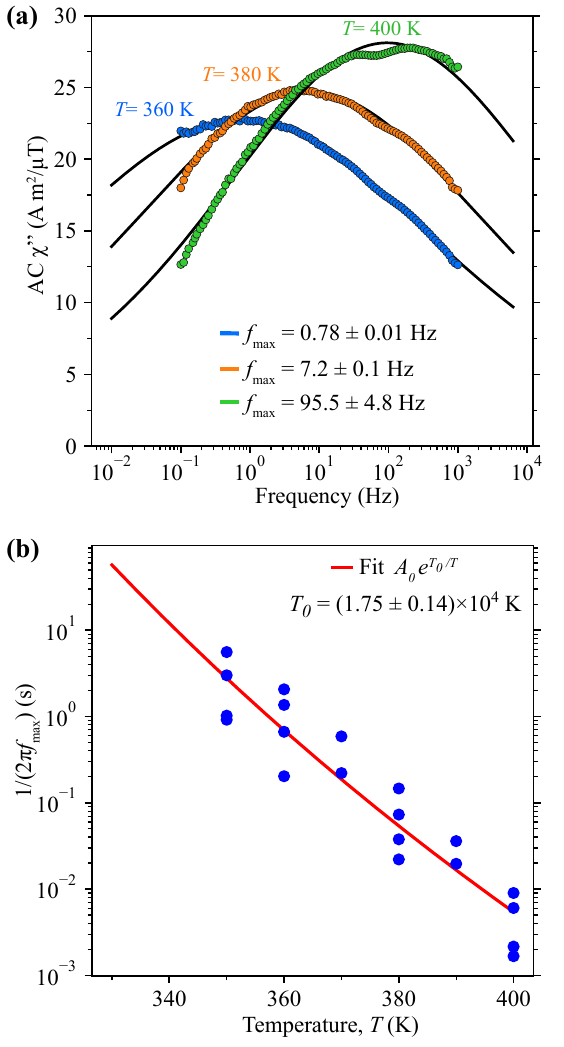} 
    \caption{\textbf{Temperature-dependent AC susceptibility.} (a) MPMS measurements of the imaginary part of the AC magnetic susceptibility versus frequency, taken with an ensemble of 30 nm SPIONs, at three different temperatures. The curves are fit to an empirical function (Eq.~A.3 of Ref.~\cite{TOP2019}) to find peak frequencies $f_{\rm max}$. (b) SPION modal relaxation time $1/(2\pi f_{\rm max})$ versus temperature, as determined from MPMS AC susceptibility measurements. The data are fit to the function $A_0 e^{T_0/T}$, with $T_0 = (1.75 \pm 0.14)\times 10^4$ K.
    }
    \label{fig:ac_suscept}
\end{figure}

The MPMS SQUID magnetometer was used to measure the AC magnetic susceptibility of a sample of the same type of SPIONs (Ocean Nanotech, Fe$_3$O$_4$, 30 nm diameter, oleic acid coated) as a function of the frequency of an applied magnetic field. For an idealized system, the imaginary part of this AC magnetic susceptibility exhibits a peak at a frequency $f_{\rm max}$ that is related to the modal relaxation time $\tau_N$ of the system by $\tau_N = 1/(2\pi f_{\rm max})$ \cite{TOP2019}. These measurements were made on a sample of SPIONs at temperatures $T = 360$ K, $380$ K, and $400$ K, and the results are shown in Figure \ref{fig:ac_suscept}(a). A shift toward higher $f_{\rm max}$ (shorter characteristic relaxation time) is observed with increasing temperature. Figure \ref{fig:ac_suscept}(b) plots measured values of $1/(2\pi f_{\rm max})$ against the temperature at which they were measured, for SPION samples of the same type. The data are fit to an exponential function, $1/(2\pi f_{\rm max})=A_0 e^{T_0/T}$, that captures the expected temperature dependence of $\tau_N$, see Eq.~\eqref{eq:neel}. 

The SPIONs imaged by diamond magnetic microscopy overwhelmingly exhibited relaxation on timescales faster than $10^{-1}$ s. The fit in Figure \ref{fig:ac_suscept}(b) indicates that this corresponds to a temperature of ${\sim}370$ K or higher. To determine a more precise effective temperature, we analyzed the MPMS AC susceptibility distribution in more detail.

The bulk AC susceptibility data is approximately log-normally distributed. The peak of the function fit to the 400 K data in \ref{fig:ac_suscept}(a) is $f_{\rm max} \approx 80~{\rm Hz}$, which corresponds to a modal value of $\tau_N \approx 2~{\rm ms}$. Taking the holding-field dependence $\tau_N = \tau_{N,0} \exp(-B_{\rm hold}/B_N)$, as observed in Fig.~\ref{fig:relax ensemble}(b) with $B_N\approx0.7~{\rm mT}$, this translates to a modal value $\tau_N \approx 0.1~{\rm ms}$ at $B_{\rm hold}=2~{\rm mT}$. This peak value of $\tau_N$, together with a $\sim 6$-decade FWHM, give a log-normal distribution that approximately matches the experimental histogram data in Fig.~\ref{fig:relax ensemble}(b) ($N = 83$ for $\tau_N < 0.06~{\rm s}$, $N = 11$ for $0.06~{\rm s}< \tau_N < 10~{\rm s}$, and $N = 3$ for $\tau_N > 10~{\rm s}$). 

\begin{figure}[htb]
    \includegraphics[width=\textwidth]{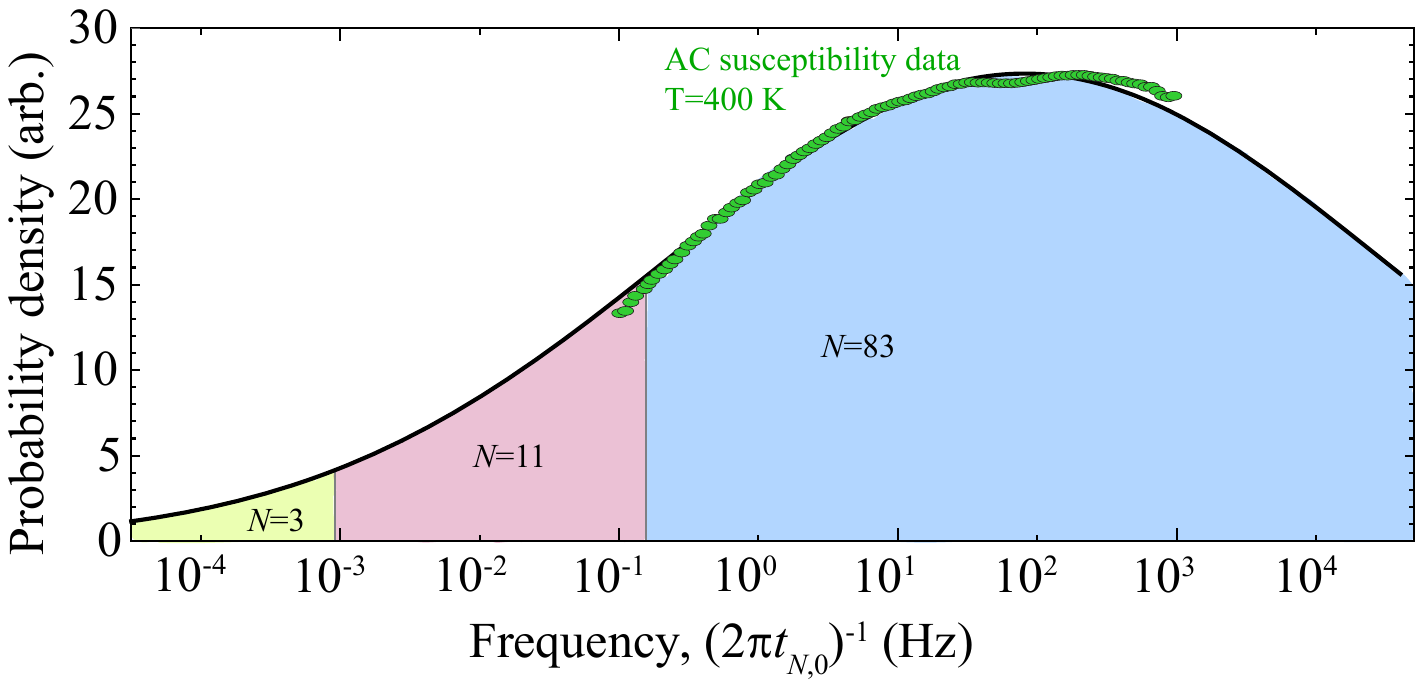} 
    \caption{\textbf{Relaxation times and AC susceptibiliy.} A log-normal function (black curve) approximates the distribution of SPION relaxation times shown in Fig.~\ref{fig:relax ensemble}, after extrapolating to $B_{\rm hold}=0$. AC susceptibility data (green dots) taken at 400 K, also presented in Fig.~\ref{fig:ac_suscept}(a), is shown alongside.}
    \label{fig:logNormal}
\end{figure}

Figure \ref{fig:logNormal} shows this estimated SPION relaxation distribution (extrapolated to $B_{\rm hold}=0$) as a function of frequency. The distribution is plotted together with the $400~{\rm K}$ AC susceptibility data, exhibiting a good match. As discussed in the main text, we consider local laser heating of the SPIONs as a possible explanation \cite{SAN2018} for the high effective SPION temperature.

\section{Relaxation measurements}
\label{app:relax_curves}
In the time-resolved magnetic imaging protocol described in Sec.~\ref{sec:relaxation measurement method}, a magnetic image is generated every 62 ms, beginning 124 ms after the polarizing field is switched off. For a given isolated SPION, the magnetic image at each time point is fit as described in \ref{app:dipoleFits}. The resulting fit parameters $\vec{m}(t)$ provide a measurement of the SPION's magnetic moment components at each time point. To quantify the SPION's Néel relaxation, we form the quantity $\Delta m(t) \equiv |\vec{m}(t) - \vec{m}(t\rightarrow \infty)|$. Here $\vec{m}(t\rightarrow \infty)$ is the SPION's fitted magnetic moment obtained in a separate measurement where the SPIONs are imaged over several hours at a static applied field with magnitude equal to $B_{\rm hold}$. The quantity $\Delta m(t)$ thus quantifies the magnitude of the vector difference between the SPION's magnetic moment at a point in time following the polarizing field relative to the magnetic moment it relaxes to after a long time. It is necessary because we perform all SPION relaxation measurements at a non-zero holding field, where the SPION magnetic moment can be substantial. Future iterations of our method may avoid this extra processing step by using much smaller holding fields, where the long-term magnetic moment approaches zero.

To find the SPION's relaxation time, $\Delta m(t)$ is fit to a decaying exponential function, $\Delta m_0\exp(-t/\tau_N) + \Delta m_{\rm offset}$. SPIONs with a best-fit $\tau_N$ longer than 20 s, or with a fit standard error for $\tau_N$ larger than 500 s, were classified as outliers. There were found to be 3 such outliers. Their relaxation curves were fit separately to an exponential decay without an offset in order to ensure that the time constant captured the slow-relaxing behavior. SPIONs with a best-fit $\tau_N\lesssim60~{\rm ms}$ were classified as short-relaxation-time outliers, denoted as having relaxed too quickly to resolve in this imaging setup. We observe that 83 of the single SPIONs studied here fall into this category.

\section{Magnetization curves for all 101 SPIONs} \label{app:data}
Figures~\ref{fig:app1}-\ref{fig:app5} shows plots of the fitted values for $m_x$, $m_y$, and $m_z$ versus $B_0$ for each of the 101 SPIONs studied in Fig.~\ref{fig:avg_lang} of the main text. The $m_x(B_0)$ curves are fit to a Langevin function, see Eq.~\eqref{eq:langevin} of the main text. There are three features (numbered 109, 117, and 140) that are designated as outliers due to having fit results $B_c > 7~{\rm mT}$ and $m_{\rm sat} > 1.2~{\rm A{\cdot}nm^2}$. These outliers exhibit little-to-no saturation at the highest applied fields, so their Langevin fits are ambiguous. 

\section{Relaxation curves for 93 SPIONs.}
Figures~\ref{fig:appRel1}-\ref{fig:appRel3} show plots of $\Delta m(t)$ for 93 of the SPIONs in the field of view.

\begin{figure*}[htb!]
    \centering
    \includegraphics[width=\textwidth]{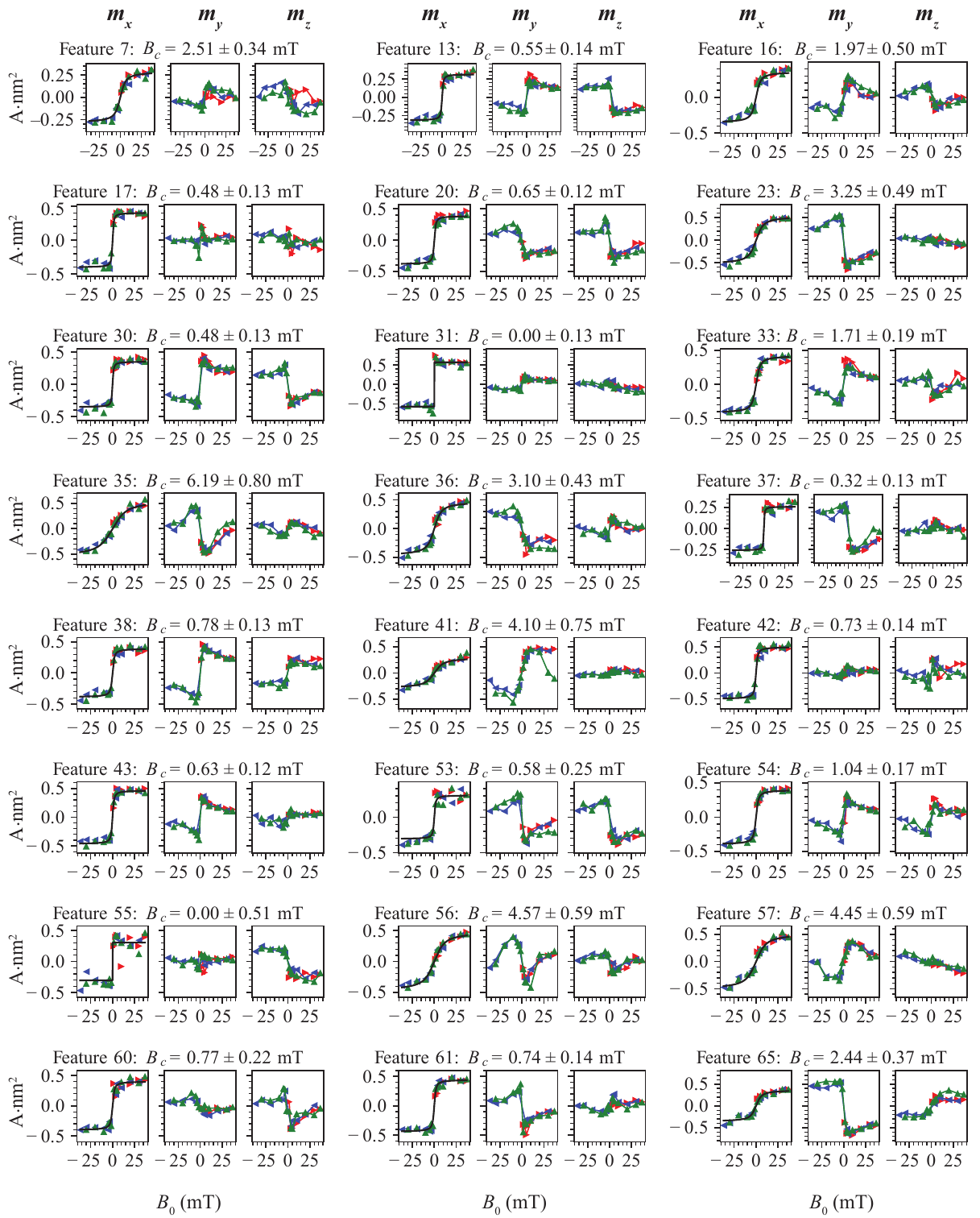}
    \caption{\textbf{$m$ vs. $B_0$ }}
    \label{fig:app1}
\end{figure*}

\begin{figure*}[htb!]
    \centering
    \includegraphics[width=\textwidth]{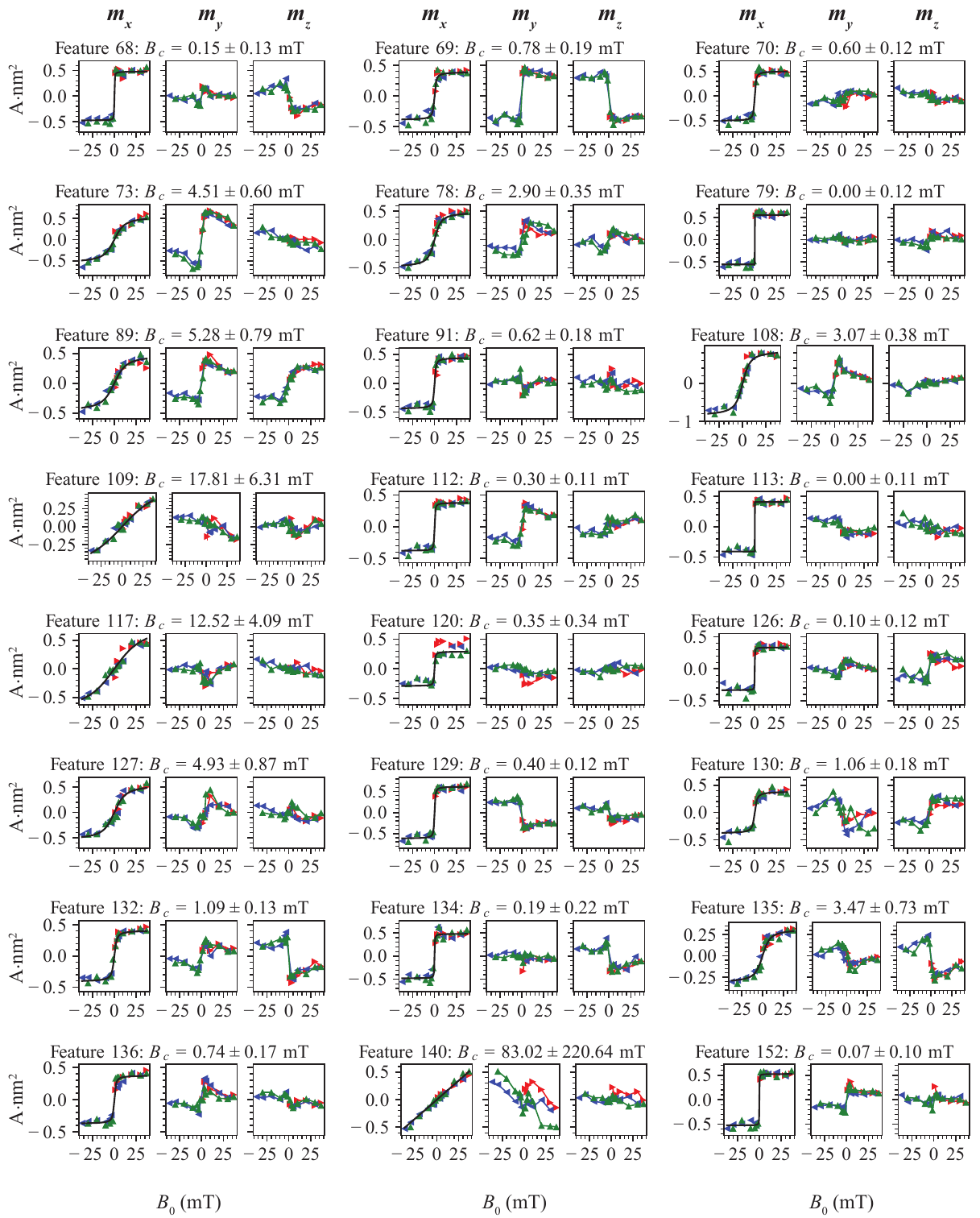}
    \caption{\textbf{$m$ vs. $B_0$ }}
    \label{fig:app2}
\end{figure*}

\begin{figure*}[htb!]
    \centering
    \includegraphics[width=\textwidth]{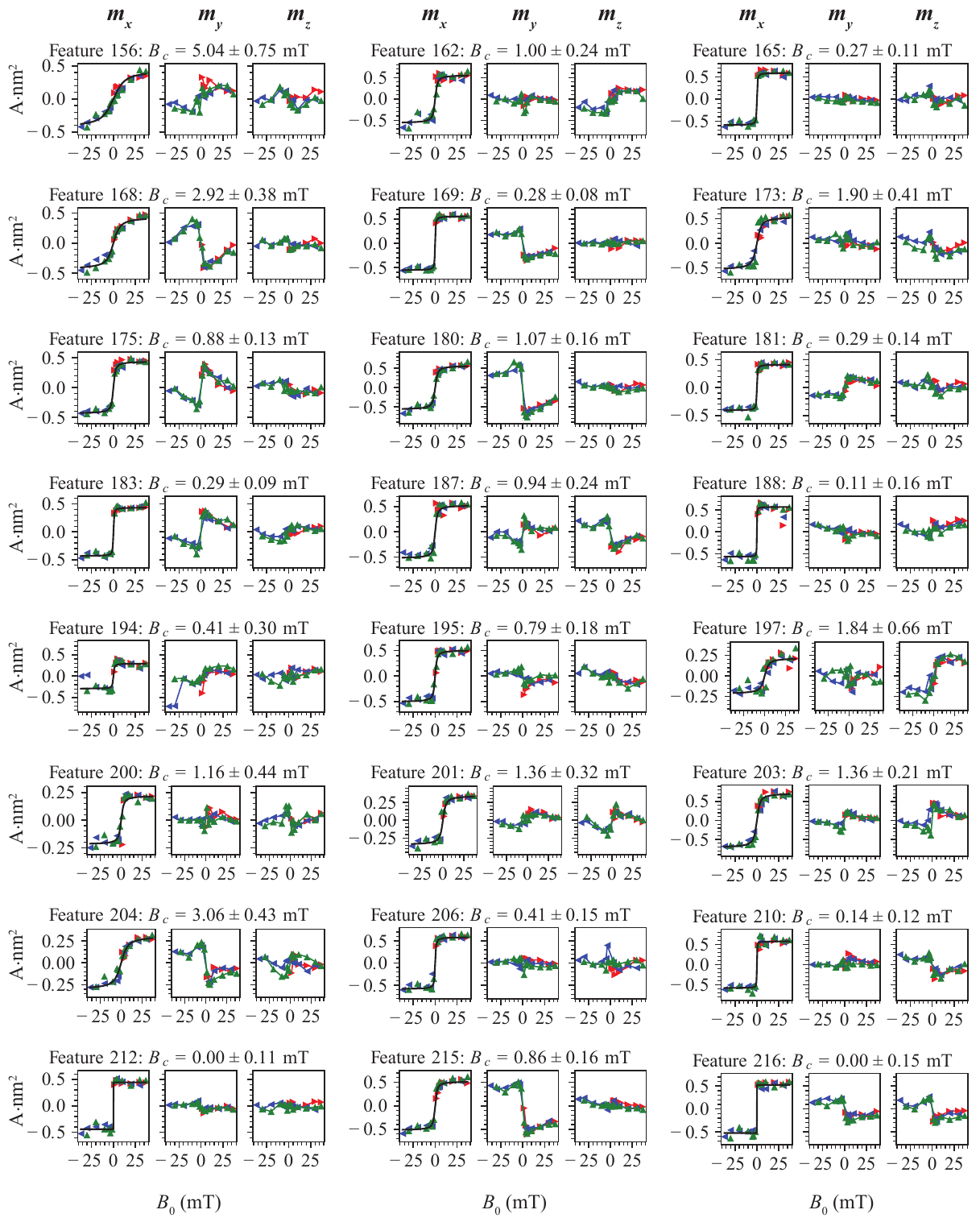}
    \caption{\textbf{$m$ vs. $B_0$ }}
    \label{fig:app3}
\end{figure*}

\begin{figure*}[htb!]
    \centering
    \includegraphics[width=\textwidth]{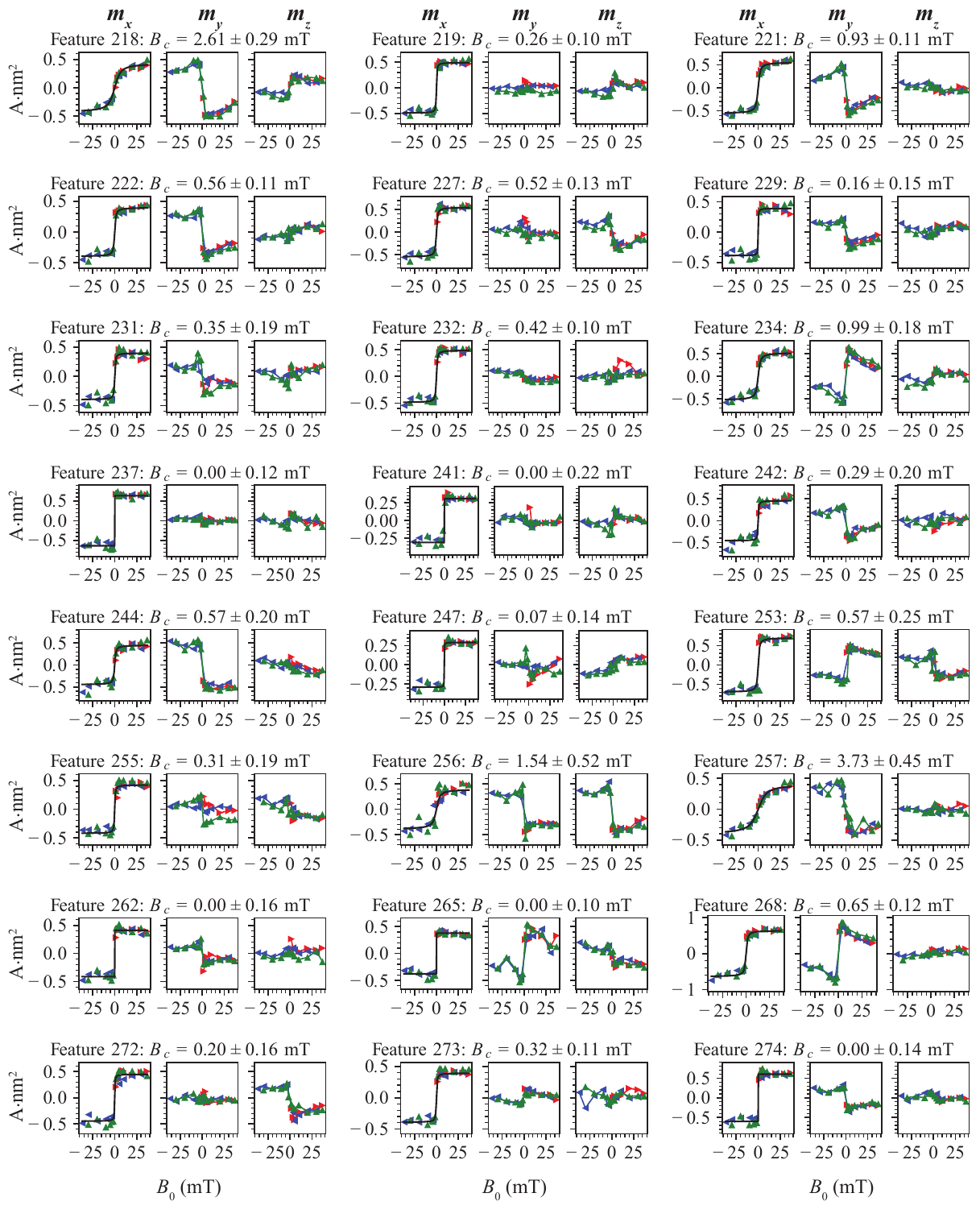}
    \caption{\textbf{$m$ vs. $B_0$ }}
    \label{fig:app4}
\end{figure*}

\begin{figure*}[htb!]
    \centering
    \includegraphics[width=\textwidth]{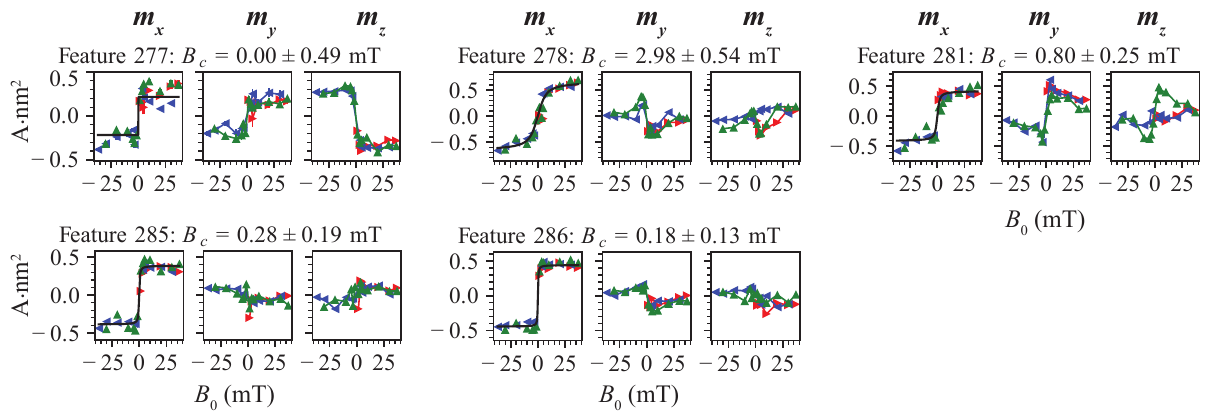}
    \caption{\textbf{$m$ vs. $B_0$ }}
    \label{fig:app5}
\end{figure*}

\begin{figure*}[htb!]
    \centering
    \includegraphics[width=\textwidth]{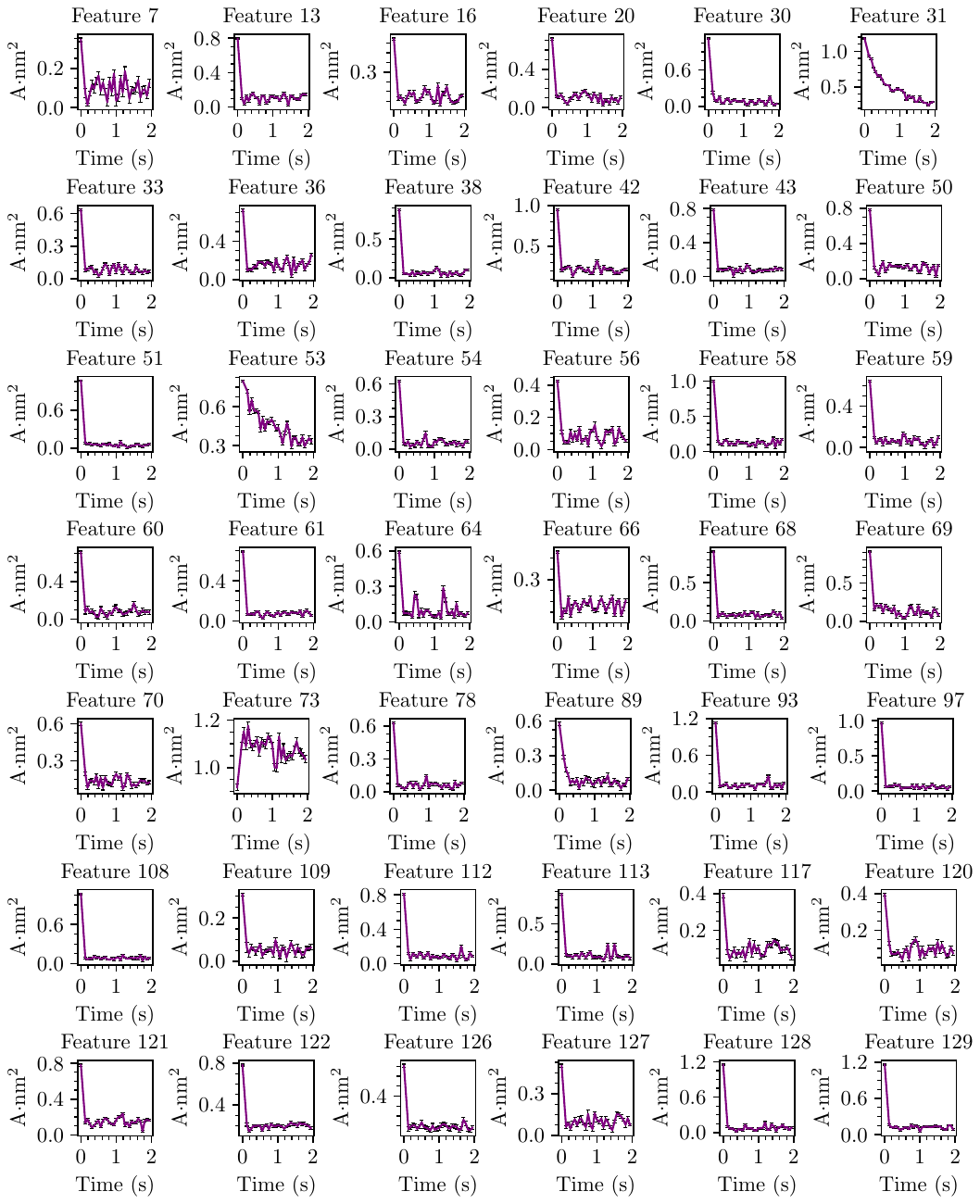}
    \caption{\textbf{Relaxation curves }}
    \label{fig:appRel1}
\end{figure*}

\begin{figure*}[htb!]
    \centering
    \includegraphics[width=\textwidth]{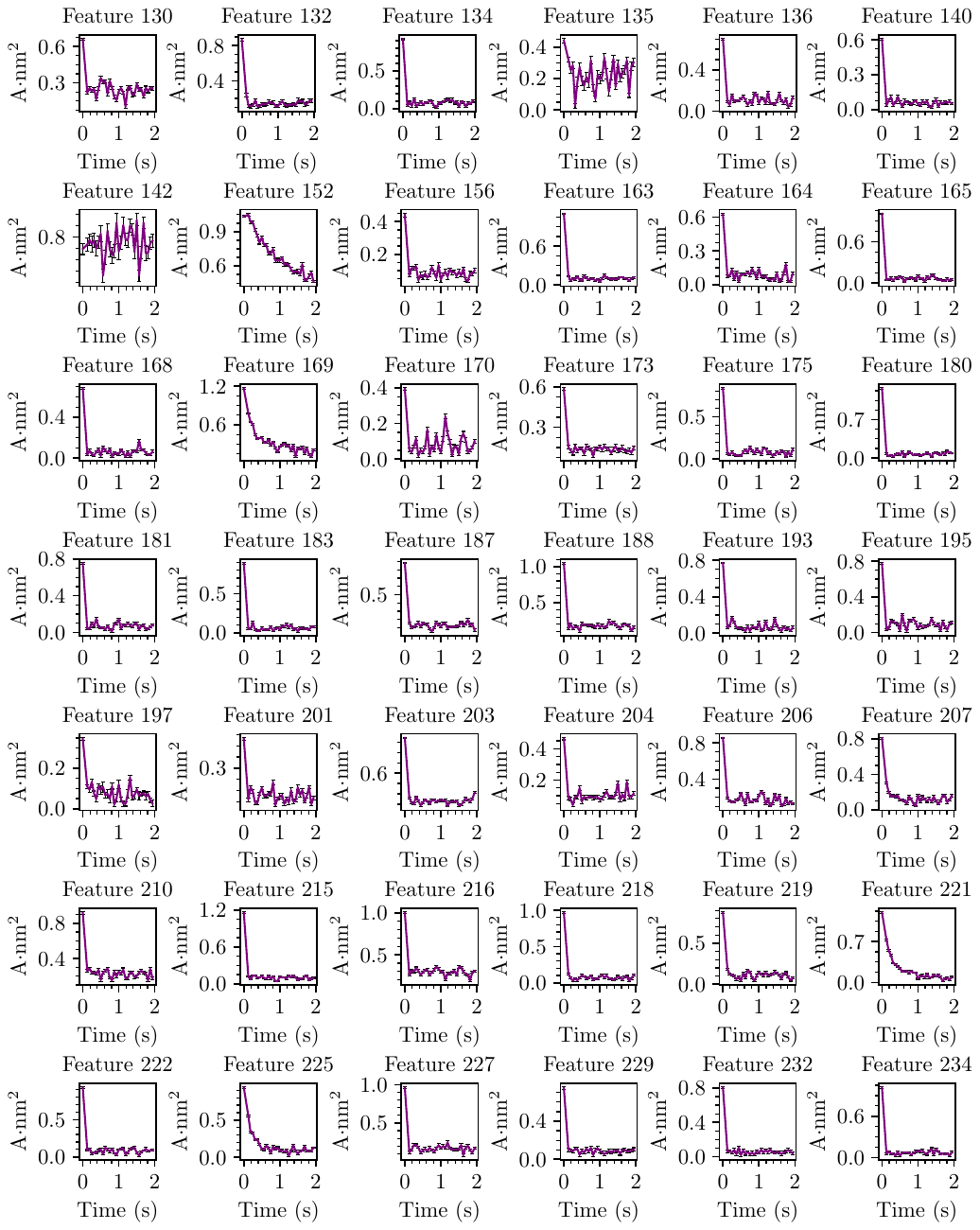}
    \caption{\textbf{Relaxation curves }}
    \label{fig:appRel2}
\end{figure*}

\begin{figure*}[htb!]
    \centering
    \includegraphics[width=\textwidth]{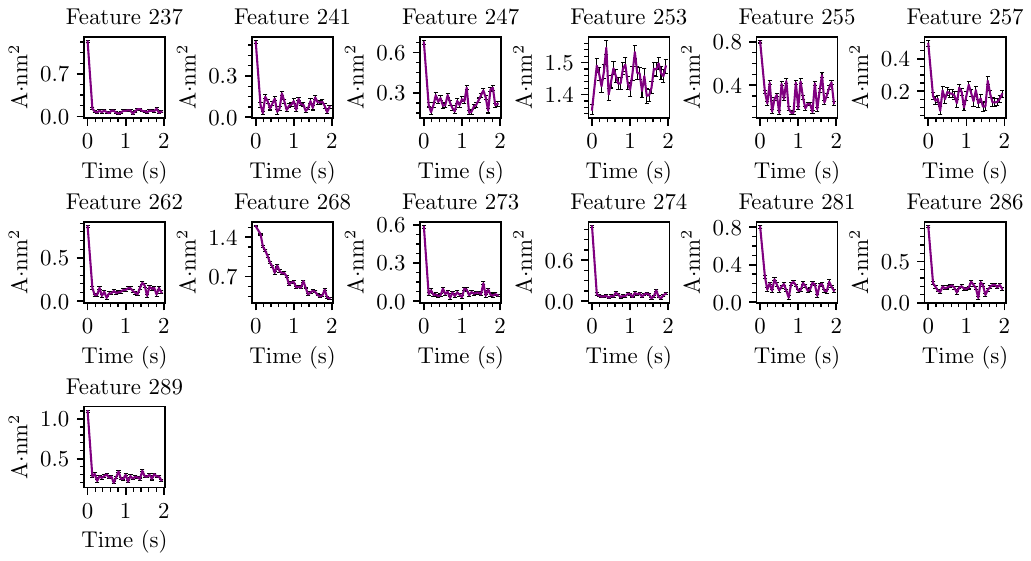}
    \caption{\textbf{Relaxation curves }}
    \label{fig:appRel3}
\end{figure*}

\clearpage

%

\end{document}